\documentclass[10pt]{article}
\usepackage[OE]{express}
\usepackage{color}
\usepackage[para,online,flushleft]{threeparttable}
\usepackage{amsmath}
\usepackage{amsfonts}
\usepackage{amssymb}
\usepackage{graphicx}
\usepackage{amsbsy}
\usepackage{multirow}
\usepackage{float}
\usepackage{caption}

\begin{document}

\title{Benchmarking five numerical simulation techniques for computing resonance wavelengths and quality factors in photonic crystal membrane line defect cavities}

\author{Jakob Rosenkrantz de Lasson,\authormark{1} Lars Hagedorn Frandsen,\authormark{1} Philipp Gutsche,\authormark{2} Sven Burger,\authormark{2} Oleksiy S. Kim,\authormark{3} Olav Breinbjerg,\authormark{3} Aliaksandra Ivinskaya,\authormark{4} Fengwen Wang,\authormark{5} Ole Sigmund,\authormark{5} Teppo H\"ayrynen,\authormark{1} Andrei V. Lavrinenko,\authormark{1} Jesper M\o rk\authormark{1} and Niels Gregersen\authormark{1,*}}

\address{\authormark{1}DTU Fotonik, Department of Photonics Engineering, Technical University of Denmark, \O rsteds Plads, Building 343, DK-2800 Kongens Lyngby, Denmark\\
	\authormark{2}Zuse Institute Berlin (ZIB), Takustra\ss e 7, D-14195 Berlin, Germany\\
	\authormark{3}DTU Elektro, Department of Electrical Engineering, Technical University of Denmark, \O rsteds Plads, Building 348, DK-2800 Kongens Lyngby, Denmark\\
	\authormark{4}ITMO University, Birzhevaja line, 14, 199034 St. Petersburg, Russia\\
	\authormark{5}DTU Mekanik, Department of Mechanical Engineering, Technical University of Denmark, Nils Koppels All{\'e}, Building 404, DK-2800 Kongens Lyngby, Denmark}

\email{\authormark{*}ngre@fotonik.dtu.dk}



\begin{abstract}
We present numerical studies of two photonic crystal membrane microcavities, a short line-defect cavity with relatively low quality ($Q$) factor and a longer cavity with high $Q$. We use five state-of-the-art numerical simulation techniques to compute the cavity $Q$ factor and the resonance wavelength $\lambda$ for the fundamental cavity mode in both structures. For each method, the relevant computational parameters are systematically varied to estimate the computational uncertainty. We show that some methods are more suitable than others for treating these challenging geometries.
\end{abstract}

\ocis{(000.3860) Mathematical methods in physics; (000.4430) Numerical approximation and analysis; (050.1755) Computational electromagnetic methods; (050.5298) Photonic crystals;  (140.3945) Microcavities; (140.4780) Optical resonators.} 


\section{Introduction}
Photonic crystals (PhCs) are periodic semiconductor material systems, which may be used for confining, guiding and generally controlling light in wavelength-scale structures~\cite{Joannopoulos2008}. As an example, spontaneous emission control from quantum dots embedded inside three-dimensional (3D) inverse opal PhC cavities has been demonstrated~\cite{Lodahl2004}. 
3D PhCs possess the advantage of exhibiting complete band gaps, but are, in turn, complicated to fabricate, and PhC membranes that exhibit partial band gaps in the plane and index confinement in the vertical direction have been proposed as a practical alternative~\cite{Busch2007}. 
With their planar nature, PhC membranes constitute a viable platform for miniaturized and integrated optical circuits~\cite{Krauss2003}, where cavities and line defect waveguides may play a key role in realizing classical functionalities such as compact optical switches, lasers, and amplifiers~\cite{Notomi2011} as well as quantum functionalities such as highly efficient sources of single photons emitted into a waveguide~\cite{MangaRao2007,Lecamp2007b}. 
Both point and line defects in the PhC membrane in-plane lattice may create cavities, and such structures are now routinely being investigated, e.g. for cavity quantum electrodynamics experiments with single quantum emitters embedded~\cite{Lodahl2015}. PhC cavities, and resonators more generally, are typically discussed in terms of their \textit{modes} that, for example, show up as peaks in a scattering spectrum. In practical cavities, these peaks are broadened; the cavity mode dissipates power in the PhC and radiates power out of the PhC, as quantified by the corresponding quality ($Q$) factor proportional to the photon lifetime~\cite{Jackson1998Chap8}. 

By leaving out a row of holes in an otherwise perfect PhC membrane lattice, a line defect is created in which light may be guided, potentially at dramatically lowered speeds~\cite{Baba2008}. If the waveguide is terminated in both ends, the finite-length waveguide forms an L$n$ cavity, where $n$ denotes the length of the cavity in terms of filled holes. Such L$n$ cavities support spectrally discrete modes, in which light may be confined for extended periods of time. These modes may give rise to enhanced light-matter interactions~\cite{Lalanne2008}, cavity quantum electrodynamics effects~\cite{Hennessy2007}, generation and on-chip routing of single photons~\cite{Yao2010}, and slow-light lasers~\cite{Xue2016}. In all cases, the cavity resonance wavelength $\lambda$ and the $Q$ factor are key parameters in the design of PhC cavities~\cite{Lalanne2008}.

Calculation of the $Q$ factor of high-$Q$ cavity modes generally represents a significant computational challenge. Numerical uncertainty, in this context defined as the variation in the computed quantity under study when relevant computational parameters such as the discretization step are varied, on the order of 10-20 \% is not unusual~\cite{Gregersen2010} even in structures where rotational or translational symmetry can be exploited to reduce the dimensionality of the problem. While the PhC L$n$ cavity does feature exploitable mirror symmetries, it still represents a full 3D problem, where determination of the $Q$ factor requires an accurate description of the geometrical variations along all three axes. When using a numerical method which expands the geometry on a grid, the required number of grid points scales with the size to the power of three leading to typical memory requirements on the order of tens of GB. Additionally, since the $Q$ factor describes leakage of light from the cavity into open space, and since most numerical simulation techniques used in photonics operate on a finite-sized computational domain, the implementation of an absorbing boundary condition~\cite{Berenger1994}, which mimics an open structure, is often required. The finite size of the computational domain as well as the specific implementation of the absorbing boundary ideally should not influence computed results, but often does so~\cite{DeLasson2015a}. Thus, when investigating the accuracy in calculation of $\lambda$ and $Q$, it is necessary to study not only the \textit{spatial resolution} of the geometry and the unknown vector quantity of interest, for instance the electric field, but also the influence from the \textit{size of the computational domain}. 

In this work, we study the fundamental cavity mode of PhC membrane L$n$ cavities, for which we determine the resonance wavelength $\lambda$ and the cavity $Q$ factor for two cavity lengths using five well-established numerical simulation techniques, namely the finite-difference time-domain (FDTD) technique, the finite-difference frequency-domain (FDFD) technique, the finite elements method (FEM), the {aperiodic} Fourier modal method ({aFMM}) and the surface integral equation (SIE) approach. A similar study of PhC nanobeam cavities using four numerical techniques was reported in~\cite{Maes2013}, where cavity frequencies and $Q$ factors were investigated as function of physical parameters. Qualitative agreement between the four methods was found, but with quantitative discrepancies in some cases being as large as an order of magnitude, and no computational uncertainty estimates were given. {Similar discrepancies when comparing methods were reported \cite{Minkov2013} in a study of six PhC cavity designs.} Thorough convergence checks of calculations of PhC cavity mode wavelengths and $Q$ factors using FDTD~\cite{Bordas2007}, FDFD~\cite{Ivinskaya2011} and FEM~\cite{Burger2010} have been reported separately, and the accuracy of the FDTD method in computing whispering gallery modes was assessed in~\cite{Boriskin2008}. Additionally, the convergence of the $Q$ factor computed for a cavity formed of subwavelength gratings~\cite{Taghizadeh2016} and of the spontaneous emission $\beta$ factor in PhC waveguides~\cite{Lecamp2007a} has been investigated using the {aFMM}. In these works, except for~\cite{Bordas2007}, however, only the spatial resolution was studied, while the influence of the domain size was not considered.

In this article, we consider two PhC structures, an L5 and an L9 cavity, and study the convergence of $\lambda$ and $Q$ as function of relevant computational parameters including the spatial resolution as well as the size of the computational domain when relevant. We provide quantitative estimates of the associated numerical uncertainty and compare the results obtained by the five methods. The computational cost in terms of CPU cores, computation time and memory requirement is compared for the different methods, and, finally, we discuss the possible origin of discrepancies larger than the numerical uncertainty in the results. 

The article is organized as follows: We first introduce the PhC membrane cavities under study in Section~\ref{Sec:PhCStructure} and discuss the basic physics of these structures. In Section~\ref{Sec:NumericalMethods}, we briefly review the five numerical methods employed. In Section~\ref{Sec:L5Results}, we present numerical results from all five methods for a smaller L5 PhC membrane cavity, and in Section~\ref{Sec:L9Results} we present similar results obtained for a larger L9 structure. The computational requirements are discussed in Section~\ref{Sec:CompRes}. The results for the two structures and from all the methods are compared in Section~\ref{Sec:Discussion} with particular emphasis on the associated numerical error. We provide a conclusion and outlook in Section~\ref{Sec:Conclusion} and detailed geometrical data and all results in a table format in Appendix~A.

\section{Photonic crystal membrane L\textit{n} cavity: Structure and physics} \label{Sec:PhCStructure}
\begin{figure}
\centering
\includegraphics[width=8 cm]{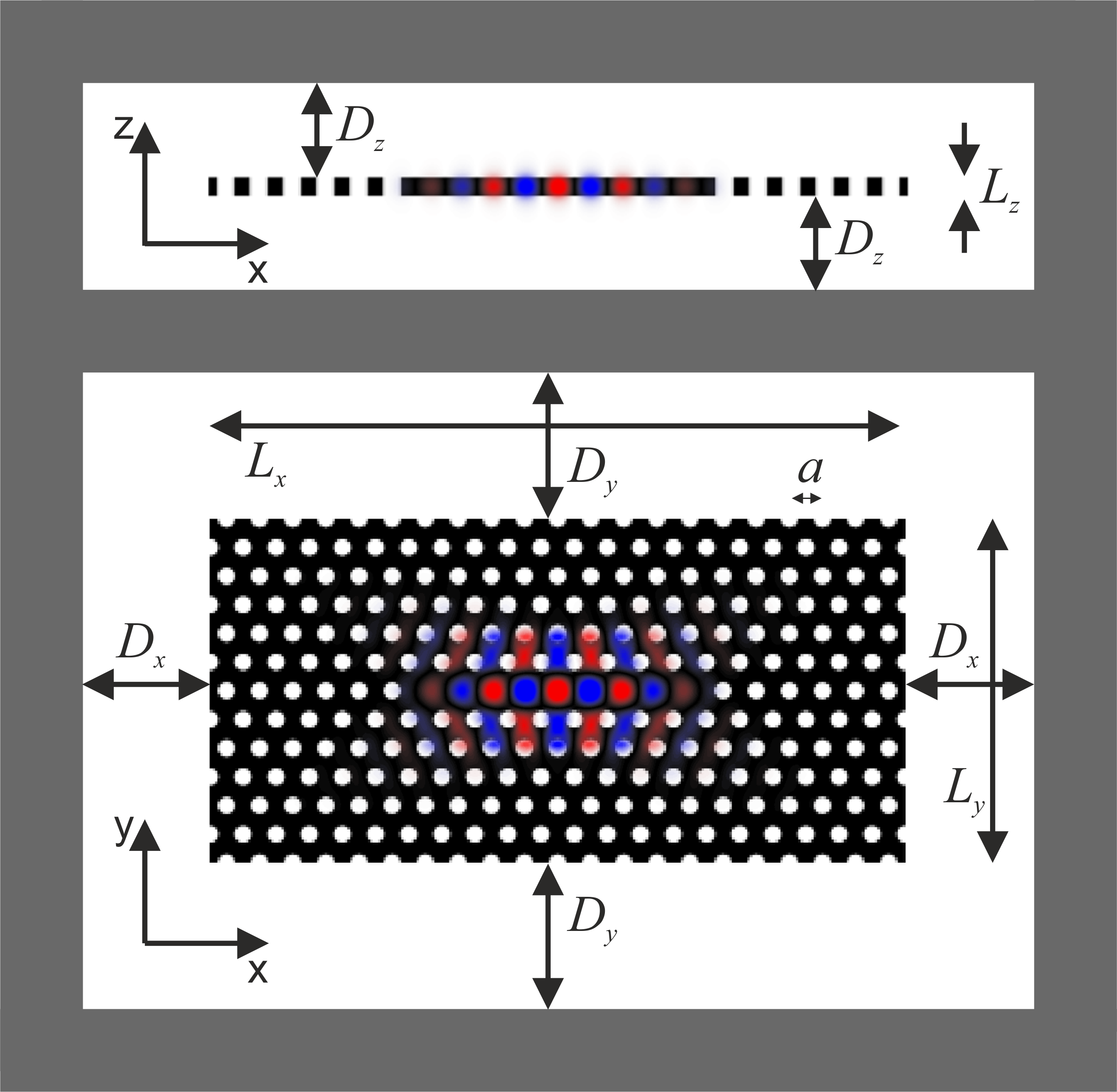}
\caption{PhC membrane geometry (black) and electric field $E_y$ profile for the M1 mode in the L9 cavity in the $y = 0$ (top) and the $z = 0$ (bottom) plane corresponding to the center of the cavity and membrane, respectively. The parameters $D_x$, $D_y$ and $D_z$ are the computational distances from the membrane to the perfectly matched layer (dark grey).}
\label{Fig:L9_ModeProfile}
\end{figure}

We consider a finite-size semiconductor PhC membrane of length $L_x$, width $L_y$, and height $L_z$, see Fig. \ref{Fig:L9_ModeProfile}. The structure is perforated by an in-plane ($xy$) triangular lattice of circular air cylinders and is surrounded by free space on all sides. The boundary of the structure is chosen such that the boundary cylinders are half circles as shown in Fig. \ref{Fig:L9_ModeProfile}. For the semiconductor material, we choose indium phosphide (InP), with refractive index $n_{\mathrm{InP}}$, that is also used in experiments with PhC membranes~\cite{Xue2016}. {While the original $L$3 design \cite{Akahane2003} featured shifted holes at the ends of the defect waveguide to increase the $Q$ factor, we will consider cavities without shifted holes corresponding to the PhC laser cavity designs of Ref.~\cite{Xue2016}. We do not expect that alternative geometries will have a noteworthy impact regarding our conclusions, although they may create some challenges for the fixed mesh approaches included in the study, cf. FDTD and FDFD, since these approaches are not ideally suited for modeling intricate boundary movements.} {Furthermore, we remark that the choice of the finite-size membrane instead of an infinite membrane was based on the difficulty of the SIE approach in handling structures that extend into infinity.}

\begin{table}
	\centering
	\captionsetup{font=bf}
	\caption{\textbf{Parameters of PhC membranes.}} \label{Tab:PhCParameters}
	\begin{tabular}{ccccccc}
		\hline 
		Geometry & $n_{\mathrm{InP}}$ & $a$ (nm) & $r$ (nm) & $L_x^{\mathrm{PhC}}$ ($a$) & $L_y^{\mathrm{PhC}}$ ($\sqrt{3}a$) & $L_z^{\mathrm{PhC}}$ (nm) \\
		\hline 
		L5 & 3.17 & 438 & 110 & 11 & 3 & 250 \\
		L9 & 3.17 & 438 & 110 & 21 & 6 & 250 \\
		\hline 
	\end{tabular}
\end{table}

According to MPB simulations \cite{Johnson2001}, the periodic PhC structure exhibits a partial bandgap in the plane in the frequency range from 1372 to 1594 nm, and by removing $n$ air holes centrally in the structure, an L$n$ cavity is formed. For two lengths of the cavity, $n = 5$ and $9$, we determine the wavelength and $Q$ factor of the fundamental mode in these cavities, the so-called M1 cavity mode, as function of computational parameters. Structural parameters are given in Table~\ref{Tab:PhCParameters}, with $a$ and $r$ denoting the PhC lattice constant and air hole radius, respectively. We stress that the surrounding air region is, in principle, of infinite extent.
In Fig.~\ref{Fig:L9_ModeProfile}, we show the electric field profile of the fundamental M1 mode in the L9 cavity calculated with FDTD, and the mode features an $|E_y|$ maximum at the center of the geometry. It is noted from Fig.~\ref{Fig:L9_ModeProfile} that the mode extends outside the cavity in the plane of the PhC membrane as well as perpendicular to this. For all five numerical methods, we verify qualitatively, by visual inspection, that the calculated modes have this field distribution.

For PhC L$n$ cavities, the limiting case of $n \rightarrow \infty$ corresponds to a so-called W1 waveguide, and it is instructive to think of the modes in the finite-$n$ cavities ($n$ = 5 and 9 here) as the discretized solutions that emerge when the W1 waveguide is terminated in both ends~\cite{Lalanne2008}. The waveguide supports a spectral continuum of propagating Bloch modes, described by a waveguide dispersion, and the lowest frequency where the guided Bloch mode exists is referred to as the cut-off or band edge. 

\begin{figure}
	\begin{center}
		\noindent
		\includegraphics[width=13.4cm]{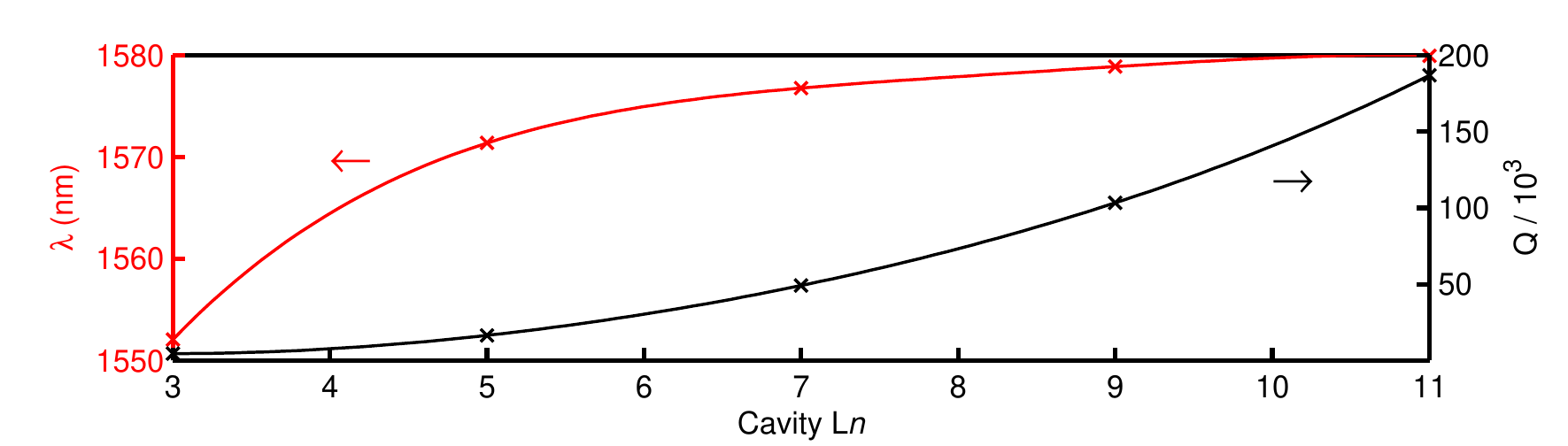}		
		\caption{Resonance wavelength (red, left axis) and $Q$ factor (black, right axis) as function of the cavity L$n$ computed using the SIE approach. In all cases, the cavity is surrounded by 6 holes as in our L9 configuration. {The solid lines are curve fits to guide the eye.}}
		\label{Fig:L_var}
	\end{center}
\end{figure}

In an L$n$ cavity, the mode existing at the lowest frequency is the fundamental mode, here called M1, and this frequency will always be larger than the band edge frequency due to the confinement of the cavity modes along the cavity (or waveguide) axis. 
Figure~\ref{Fig:L_var} shows the dependence of the resonance wavelength and $Q$ versus the order $n$ of the cavity L$n$. The shortest cavities feature the shortest wavelengths, whereas for increasing cavity length (larger values of $n$), the fundamental cavity mode wavelength will approach the band edge asymptotically~\cite{Okano2010}. The other important figure of merit is the $Q$ factor, and this quantity increases with the length of the cavity~\cite{Okano2010}. If we again consider the cavity mode as a result of spatially confining the propagating waveguide mode, the finite $Q$ factor stems from scattering of this propagating mode into other modes (for example, the PhC membrane radiation modes) as it reaches the end of the cavity. The relative importance of these scattering losses decreases as the cavity becomes longer due to the combination of increasing cavity length and increasing group index $n_g$, the latter being due to the shift of the resonance wavelength towards the bandgap~\cite{Sauvan2005a} as the cavity length increases. 

In this work, we focus on the convergence of $\lambda$ and $Q$ for the L$5$ and L$9$ configurations described in Table \ref{Tab:PhCParameters}. In the first and third rows in Table~\ref{Tab:ResultsSummary}, we collect the best estimates for the M1 wavelength as calculated with the five numerical methods to be introduced in Section~\ref{Sec:NumericalMethods}. It is apparent that the longest cavity gives rise to the longest wavelength (smallest frequency), in agreement with the discussion above. In the second and fourth rows in Table~\ref{Tab:ResultsSummary}, we collect the best estimates for $Q$ of the fundamental L5 and L9 cavity modes, and these confirm the qualitative expectation, that $Q$ increases with $n$. At the same time, it is apparent that quantitative discrepancies are found, which we analyze further in Sections~\ref{Sec:L5Results} and \ref{Sec:L9Results}.

\begin{table}[t]
\centering 
\captionsetup{font=bf}
\caption{\textbf{Best estimates of M1 wavelengths and $Q$ factors for each computational method.}} \label{Tab:ResultsSummary} 
\begin{tabular}{cccccccc}
	\hline
 & \textbf{FDTD} & \textbf{FDFD} & \textbf{pFEM} & \textbf{sFEM} & \textbf{tFEM} & \textbf{{aFMM}} & \textbf{SIE} \\ \hline 
$\pmb{\lambda^{\mathrm{L5}}}$ \textbf{(nm)}  & 1570.9 & 1570.5 & 1570.5 & 1571.0 & 1572.5 & 1568.0 & 1571.5 \\
$\pmb{Q^{\mathrm{L5}}}$ & 1686.8 & 1714.9 & 1715.9 & 1711.5 & 1703.9 & 1417.1 & 1706.6 \\
$\pmb{\lambda^{\mathrm{L9}}}$ \textbf{(nm)} & 1577.7 & 1579.9 & 1577.5 & 1578.2 & 1579.5 & 1572.0 & 1578.6 \\ 
$\pmb{Q^{\mathrm{L9}}}$ & 104,389 & 101,289 & 105,595 & 104,387 & 104,561 & 13,507 & 103,931 \\
	\hline
\end{tabular}
\end{table}

\section{Numerical methods} \label{Sec:NumericalMethods}
In the following subsections, we briefly describe the five numerical methods investigated with focus on the computational approximations and the relevant parameters as well as the method used for determining the M1 mode wavelength and $Q$ factor.

A first common feature for all the methods is the exploitation of the symmetry planes of the geometry. When the origin is placed in the center of the cavities, the L5 and L9 geometries are symmetric across the $x$, $y$ and $z$ axes, and accordingly the optical mode profiles will be symmetric or anti-symmetric along these axes. Consequently, only one eighth of the full structure needs to be stored in computer memory allowing the treatment of significantly larger structures than would be possible in the absence of symmetry planes. 

Another common feature for the FDTD, FDFD, FEM and {aFMM} methods is the necessity of implementing an absorbing boundary condition due to the finite computational domain employed for these methods. For these methods, the geometry is thus surrounded by perfectly matched layers (PMLs) in all directions, except the {aFMM} where PMLs are only required on $y$ and $z$ boundaries (see Fig.~\ref{Fig:L9_ModeProfile}). For the SIE method there is no need of an artificial boundary condition to limit the computational domain.

Finally, a common feature for the FDFD method, the FEM when using eigenvalue analysis (pFEM, tFEM), the {aFMM} and the SIE approach is the calculation of a quasi-normal mode, defined as a solution to Maxwell's equations subject to an outgoing radiation boundary condition~\cite{Kristensen2014}. This choice of boundary condition renders the wave equation non-Hermitian leading to complex eigenfrequencies for the modes. Adopting the time-convention of $\exp(-\mathrm{i}\omega t)$, the imaginary part of the quasi-normal frequency is negative, $\mathrm{Im}(\omega) < 0$, and the corresponding $Q$ factor is deduced from the real and imaginary parts of the complex eigenfrequency as $Q = -\mathrm{Re}(\omega)/(2\mathrm{Im}(\omega))$~\cite{Lalanne2008}. The resonance wavelength $\lambda$ is given by $\lambda = 2\pi \mathrm{c}/\mathrm{Re}(\omega)$, where $\mathrm{c}$ is the speed of light. On the other hand, in the FDTD method the $Q$ is extracted from the decay rate, whereas for the sFEM it is derived from the width of the resonance peak.

\begin{figure}[]
	\centering
	\includegraphics[width=13 cm]{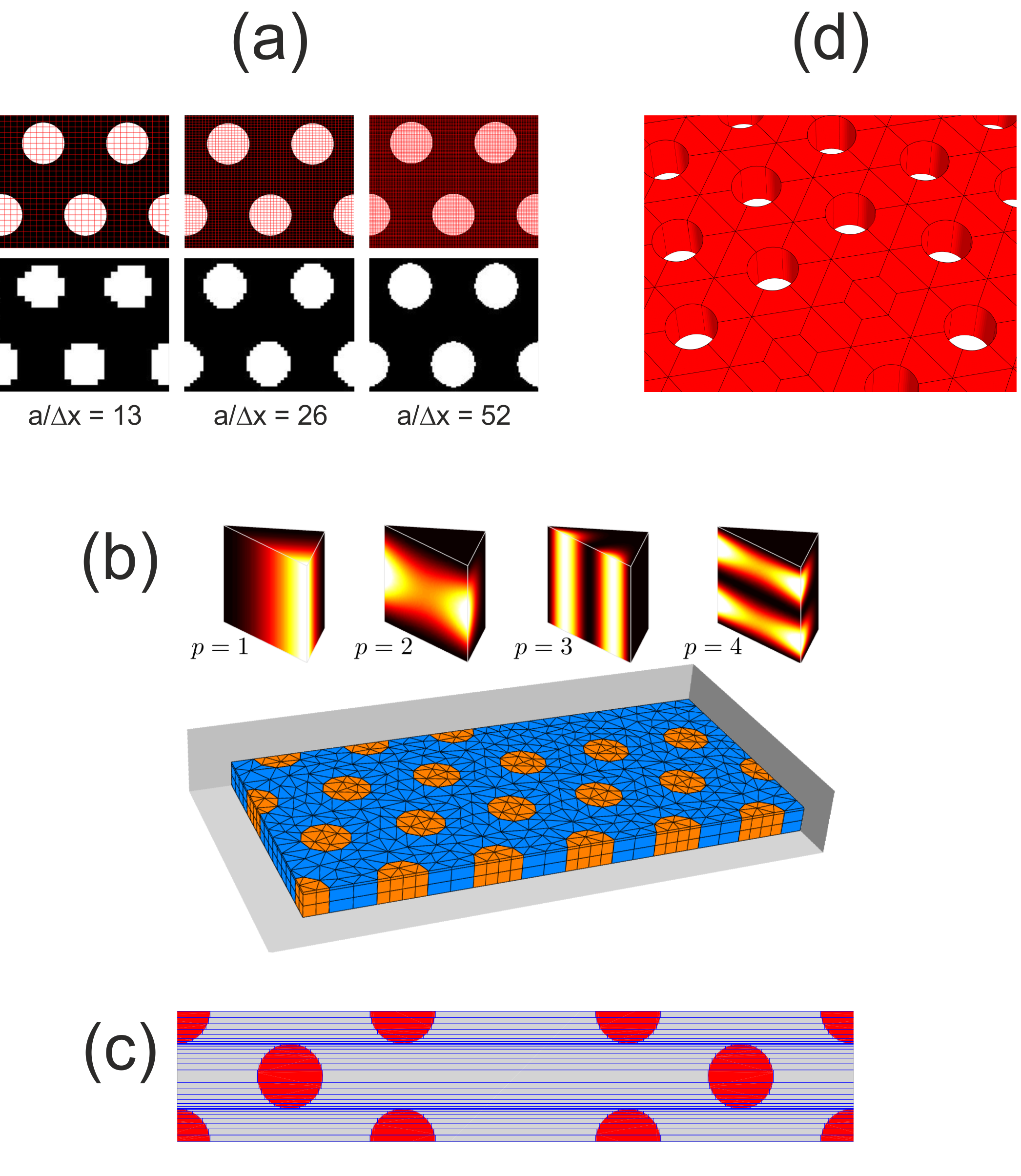}
	\caption{Discretization schemes. (a) FDTD/FDFD: {(Upper) Representative cutouts of the photonic crystal structures defined by air holes (white) in InP (black) material for different spatial mesh (red grid) resolutions. (Lower) The meshed material distributions for the corresponding cutouts.} (b) FEM: Mesh of the PhC membrane L5 geometry used in the sFEM simulations with $h$ = 130 nm. Due to symmetry planes, only one eighth of the full device is required with corresponding boundary conditions set on the grey faces. The InP PhC membrane (blue) is perforated with an in-plane triangular lattice of air cylinders (orange). The mesh consists solely of prisms in the interior and additional bricks in the exterior PML domain (not shown). Edge and face ansatz functions based on Jacobi polynomials are shown (top) for degrees $p$ = 1, 2, 3 and 4 at the top. {(c) Staircase discretization scheme employed by the {aFMM}. 33 layers are used per unit cell. (d)} SIE: Accurate representation of the geometry of the PhC membrane with higher-order quadrilateral elements.}
	\label{Fig:Disc_schemes}
\end{figure}

The methods can be categorized as either single-shot or iterative techniques. In the case of FDTD, FDFD, pFEM and tFEM, the determination of $\lambda$ and $Q$ only requires a single calculation. In contrast, the {aFMM} and the SIE require multiple calculations to find an extremum point in complex frequency space. For these methods, the proximity of a computed complex frequency to the extremum point represents an additional parameter for which a convergence criterion must be determined. This criterion should be chosen such that the uncertainty due to not having determined the exact extremum point is significantly smaller than the numerical inaccuracy due to other free parameters such as grid resolution and domain size. Finally, for the FEM using scattering analysis (sFEM), multiple calculations are required to provide a sampling of the resonance peak. Again the number of sampling points should be chosen large enough  that the error due to inaccurate sampling is not dominating over those induced by other parameter variations. 

For all the methods, a starting guess $\omega_{\text{SG}}$ ($\lambda_{\text{SG}}$) for the resonance frequency (wavelength) must be provided, even though, ideally, we would like to determine the mode characteristics without any prior knowledge. However, both the L5 and L9 structures support many longitudinal modes, and to avoid computing every confined mode, at some point a choice must be made regarding where to look along the frequency axis. In this context, we remark that the computation time, especially for the iterative methods, is dependent on the proximity of the starting guess to the mode frequency. A starting guess is generally determined by running the simulation at coarse resolution to get a rough estimate of the positions of the modes.

\subsection{Finite-difference time-domain (FDTD) method} \label{Sec:FDTD}
The finite-difference time-domain (FDTD) method is solving Maxwell's equations in differential form, and was introduced in 1966 by Yee~\cite{Yee1966} and later heralded by Taflove and co-workers~\cite{Taflove1975,Taflove2004}. Today, the method is widely used to provide unique insights into a vast number of problems in electromagnetics and photonics. FDTD is a time-domain technique and, thus, calculates the temporal evolution of electromagnetic fields propagating in finite geometries, and by applying Fourier transforms to the temporal solutions the spectral response is obtained. In the work presented here, we have used the `FDTD Solutions' software package from Lumerical~\cite{Lumerical} to model the PhC cavities.

One of the main challenges when applying the FDTD method to photonic crystals is that structures are discretized on a Cartesian grid, and thus a staircasing scheme is applied to approximate any curved material boundaries occurring, here between InP material and air as illustrated in Fig. \ref{Fig:Disc_schemes}(a). As a consequence, the sizes of the mesh elements $\Delta x$, $\Delta y$, and $\Delta z$ become important parameters for accurately modeling the geometry under study, and the smaller the size of the mesh elements, the more accurate the model. The `FDTD Solutions' software package includes the option of sub-griding the geometry to reduce the required computational resources. Here, critical parts of the structure, for example the periodically placed PhC holes surrounding the cavity, can be meshed with high spatial resolution, whereas the less critical parts, for example the homogeneous air regions surrounding the photonic crystal slab, are meshed with lower spatial resolution. 

The resonance wavelength $\lambda$ of the cavity is obtained via the Fourier transform correlation between temporal and spectral solutions providing the transmission spectrum in which peaks corresponds to the resonances of the cavity. Thus, an accurate value for $\lambda$ is obtained by having a high spectral resolution, which requires a large number of time steps $t$ to be calculated, effectively increasing the total calculation time. 
The quality factor Q of a given mode is estimated from the temporal decay of the mode using the high-Q analysis procedure provided by Lumerical Solutions and described in~\cite{Lumerical_link1}, when the cavity is excited by a spectrally narrow dipole around the mode of interest. Following the Courant-Friedrichs-Lewy stability criteria~\cite{Courant1928} for solving partial differential equations, each time step $\delta t$ is chosen to be proportional to the size of the mesh. Consequently, for a fixed total simulation time, when the mesh element size is reduced, the time step is decreased and the number of total time steps increases. A small time step, in turn, leads to a better accuracy when using the temporal decay of cavity modes to calculate high $Q$ factors. 

\subsection{Finite-difference frequency-domain (FDFD) method} \label{Sec:FDFD}
{The finite-difference frequency-domain (FDFD) method~\cite{Ivinskaya2011} solves Maxwell's equations in the frequency domain. The method is formulated on the structured grid and is conventionally regarded as a counterpart of the FDTD approach in the frequency domain. The most often used FDFD method is based on the classical Yee staggering with different offsets in the location of the field components within the discretization cell~\cite{Yee1966}. Initially, the 3D frequency-domain method was widely used in microwave cavity simulations and for analysis of waveguide modes, but now it has been extended also to photonic problems. The FDFD approach allows determination of band gap diagrams (eigenfrequencies and field maps of guided modes), as well as scattering, particularly in reflection/transmission computations.}

{In the FDFD formulation, one assigns Maxwell's differential operators as well as the permittivity $\epsilon$ and permeability $\mu$ on the discretization grid. The symmetrical matrix $\hat{\theta} = \sqrt{\mu^{-1}}\nabla^{\times}\epsilon^{-1}\nabla^{\times}\sqrt{\mu^{-1}}$ is built and used in the formulation of the algebraic eigenproblem $\hat{\theta}(\sqrt{\mu}\textbf{H}) =\omega^2(\sqrt{\mu}\textbf{H})$, where $\textbf{H}$ is the magnetic field. To solve the system of linear equations, either an iterative solver which reduces the memory requirement of the method, or the $LU$ factorization which usually requires less computation time and is more stable but also more memory consuming, is used. The eigenvalues for the quasi-normal modes are then found by the standard shift and invert procedure~\cite{Lehoucq1998}. In the frequency domain, the resonance wavelengths and the quality factors (including resolution of degenerate states and variants of their field profiles) appear straightforwardly from the real and imaginary parts of the complex eigenvalues.}

{To emulate open boundaries on a finite computational domain, perfectly matched layers (PMLs) are commonly used. We construct PMLs~\cite{Shin2012} with the coordinate-scaling method~\cite{Shyroki2007} which is very easy to implement in the frequency domain and naturally combines complex coordinate scaling for efficient dumping of oscillatory outgoing solutions and real-valued coordinate mapping to suppress evanescent waves if needed~\cite{Shyroki2010}.}

\begin{figure}[b]
	\centering
	\includegraphics[width=9 cm]{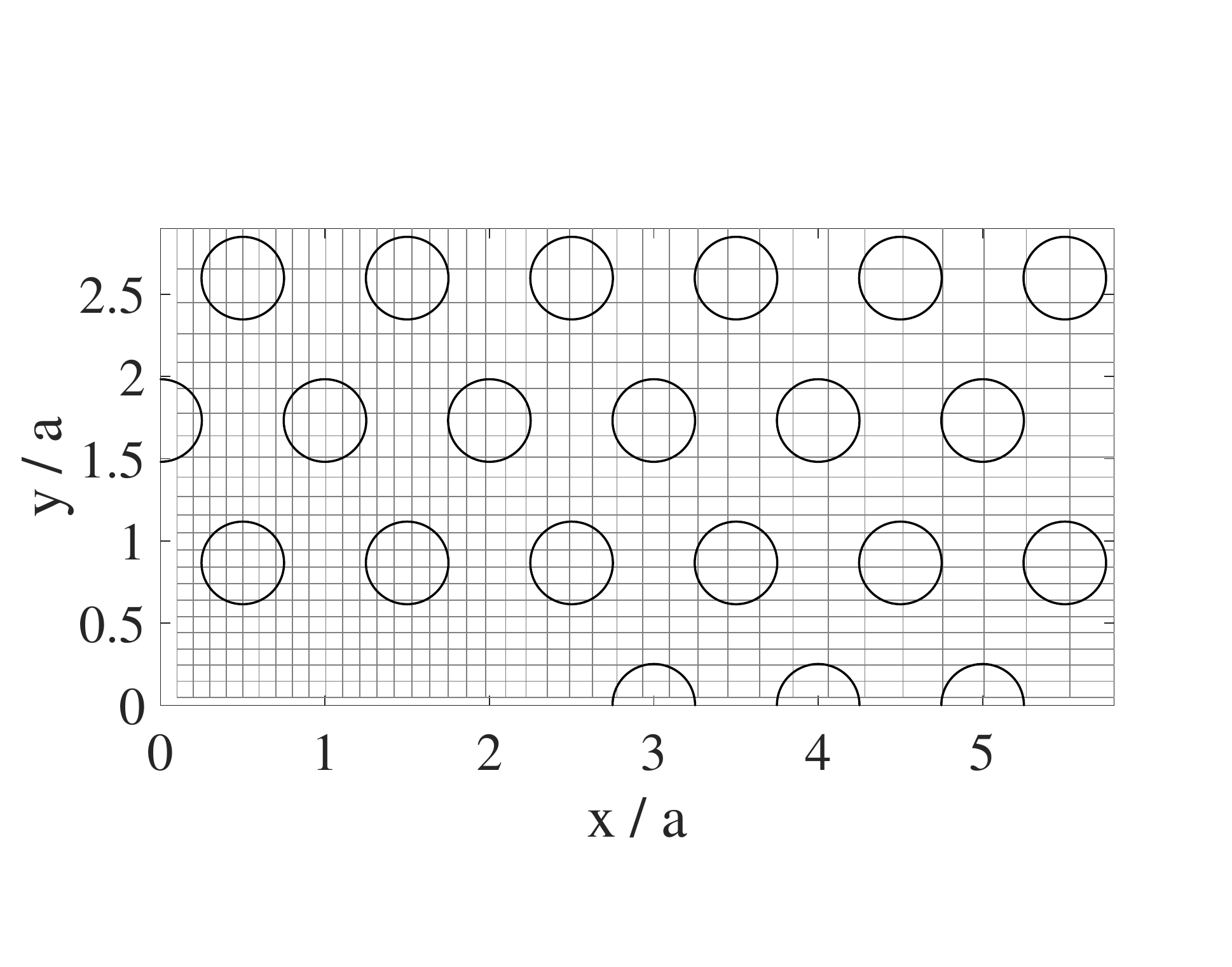}
	\caption{{Nonuniform discretization mesh in the FDFD simulations. The discretization step decreases towards the center of the cavity.}}
	\label{Fig:FDFDmesh}
\end{figure}

{In the FDFD case, similar to the FDTD method, one may construct adaptive non-uniform grids. Building an orthogonal, non-uniform grid of the continuously varying density - for example, setting a ratio 1:3 between its high- and low-density regions, cf. Fig.~\ref{Fig:FDFDmesh} - gives stable convergence for both wavelength and quality factor and significantly reduces the memory requirements. However, in contrast to the time-domain techniques requiring the time step adjusted to the smallest spatial discretization step, the frequency domain approaches are not constrained by the Courant-Friedrichs-Lewy time step limit~\cite{Courant1928}. For grid cells containing interfaces between different materials, specifically constructed dielectric tensor components can be assigned to the corresponding unit cells. In each unit cell the interface between two media is effectively flat and, knowing the normal to interface, the effective permittivity tensor can be derived similarly to standard averaging techniques for plain geometries~\cite{Shyroki2011,Lavrinenko2014}. A proper choice of an averaging technique ensures that the convergence rate remains the same as for the homogeneous regions.}
	
{The FDFD approach can be programmed with ease if high-level languages are used with convenient handling of matrix operations. Here we use a home-made realization in MATLAB.}

\subsection{Finite-element method (FEM)} \label{Sec:FEM}
The finite element method (FEM) is a numerical method for solving partial differential equations~\cite{Lavrinenko2014}. It relies on adaptive space discretization using unstructured meshes (tetrahedral, prismatoidal, triangular, etc.) as illustrated in Fig.~\ref{Fig:Disc_schemes}(b) and on expansion of the unknown solution using small sets of ansatz functions for each mesh element. The main discretization parameters are the mesh element size, $h$, and the polynomial order of the ansatz functions, $p$. It can be shown that both decreasing $h$ and increasing $p$ yields independently or simultaneously~\cite{Burger2015hp} higher accuracy of the numerical results. Typically, larger accuracy is obtained by employing higher-order polynomials. 


We use two different commercial software packages for the FEM simulations. The first is JCMsuite developed by JCMwave that includes solvers for time-harmonic (frequency-domain) Maxwell eigenvalue and scattering problems~\cite{Pomplun2007}. The second is COMSOL~\cite{Comsol52} which is discussed further below.

When computing the $Q$ factor and resonance wavelength of a cavity, one can use either an eigenvalue solver or a scattering solver~\cite{Burger2011}. Given the discretized geometry of the cavity and the material parameters, the eigenvalue solver computes the electric field distribution of the quasi-normal mode and a corresponding complex eigenfrequency $\omega$ representing a solution to the Maxwell eigenvalue problem. JCMsuite features mixed (tetrahedrons, prisms and bricks) meshes. In this study, prism meshes have been used and results obtained using the JCMsuite eigenvalue solver are labelled pFEM in the following.

In contrast to the eigenfrequency approach, solving the light scattering problem requires an (external or internal) time-harmonic source definition and computes the electromagnetic field distribution in the given geometry excited by this specific source. Here, we use a point source placed at the symmetry point of the cavity similar to FDTD calculations. For obtaining the resonance wavelength and $Q$ factor, one computes, for example, the Purcell factor~\cite{Purcell1946} as a function of frequency. This emission enhancement of the dipole source is obtained as the ratio of emitted power in the cavity structure to the power emitted in a homogeneous InP medium~\cite{Novotny2012chap8}. This yields a typical resonance spectrum with a peak for each cavity resonance. Dividing the resonance wavelength by the resonance width (full width half maximum) again gives the $Q$ factor. {In order to determine this spectrum, we use an adaptive sampling strategy in the frequency domain.} We use a fixed number $N$ = 5 of iterations with 3 frequencies for each iteration to sample the resonance peak for both the L5 and the L9 cavities. In the L5 cavity case, the resonance is relatively broad and is easily found by our search routine; an accurate sampling of the peak is obtained. In contrast, for the L9 cavity the resonance is approximately 50 times more narrow, and more iteration steps are needed to find the resonance, or, when using the same number of iteration steps as used for the L5 cavity, a lower quality sampling of the peak is obtained. This leads to a large relative deviation in the convergence studies for the L9 cavity as discussed further in Sections \ref{Sec:L9grid} and \ref{Sec:L9size}. Results obtained using the JCMsuite scattering approach are labeled sFEM.

Finally, we have used the software package COMSOL~\cite{Comsol52} to compute the $Q$ factor and resonance wavelength. Here, we use an eigenvalue analysis equivalent to the pFEM technique described above. The computational domain is discretized using isoparametric tetrahedral elements, where geometrical shape order is the same as the unknown electric field order, i.e. both represented by second order polynomials ($p$=2). Meshes in COMSOL are defined by specifying minimum and maximum element sizes $h$. The eigenvalue analysis is performed using a direct solver called 'MUMPS', and the results obtained using the COMSOL eigenvalue analysis are labeled tFEM.

{While both pFEM and tFEM solve the eigenvalue problem of Maxwell's equations, the physical space is discretized with a mesh based on different building blocks. In the case of pFEM these building blocks are solely prisms, and in the case of tFEM the mesh consists only of tetrahedrons. In both cases, the circular holes of the cavity are approximated by polygons, cf.\ Fig.~\ref{Fig:Disc_schemes}(b). The only difference in the corresponding meshes is the discretization of the volumes of the surrounding air and the InP cavity. We remark that in addition the degree of the polynomial basis spaces is varied for pFEM and tFEM independently.}

\subsection{Aperiodic Fourier modal method ({aFMM})}
For the Fourier modal method (FMM), we use an in-house developed code~\cite{DeLasson2015a}. In the FMM, the geometry is divided into layers that are uniform along the propagation $x$ axis. Eigenmodes in each layer are then determined via Fourier series expansions in the transverse $yz$ coordinates~\cite{Li1997}. The local expansions in each layer are coupled using a scattering (S) matrix routine~\cite{Li1996a}, and the structure is partitioned into a small number of periodic sections, for which the Bloch modes are determined as the eigensolutions of the supercell S matrices~\cite{Lecamp2007a}. The Bloch modes are used as an expansion basis in each periodic section, which are, finally, coupled together using a Bloch mode S matrix approach. Referring to the bottom panel in Fig.~\ref{Fig:L9_ModeProfile}, the PhC membrane structures considered here consist of five periodic sections, of which three are unique (air section, bulk PhC section, and PhC cavity section). Calculating all relevant Bloch modes thus amounts to analyzing three distinct supercells, with the air one being trivial. We exploit the $y$ and $z$ mirror symmetries to solve a reduced-size problem~\cite{Li2003}.

Along the $x$ axis, the geometry is approximated by a staircase profile {illustrated in Fig. \ref{Fig:Disc_schemes}(c) with 33 steps per unit cell, where an increasing number of steps leads to an improved representation of the true geometry.} Additionally, the Fourier series truncation order in the $y$ and $z$ directions dictates how well the geometry is approximated in these transverse directions. Truncating these series at $l_y$ ($l_z$), we include a total of $2l_y+1$ ($2l_z+1$) $y$ ($z$) dependent terms, and we later study the impact from $l_y$ and $l_z$ on resonance wavelengths and $Q$ factors in detail. Finally, the dimensions of the computational domain along the transverse directions are also parameters that we vary and analyze. 
{While originally developed for periodic structures, the FMM may be combined with an absorbing boundary condition to enable analysis of aperiodic structures~\cite{Lalanne2000,Silberstein2001}. In this work, an absorbing boundary condition is implemented at the transverse domain edges}
using a complex coordinate transformation $f_{PML}^{-1} = 1+i$~\cite{Hugonin2005a} with a fixed thickness of the PML layer of $D_{\mathrm{PML}} = \frac{3}{2}L_z^{\mathrm{PhC}}$~\cite{DeLasson2015a}. {The FMM with PMLs included is referred to as  aperiodic FMM (aFMM)}. As a perspective for future FMM work, we note that an infinite-size computational domain version of FMM has recently been developed~\cite{Hayrynen2017}, in which absorbing boundary conditions are avoided.

In the direction of the cavity axis ($x$) we have an analytic description of the propagation of both incoming and outgoing Bloch modes, and we here require to only have outgoing modes in the air regions surrounding the membrane. This condition together with an iteration of the frequency in the complex plane to find a unity eigenvalue of the so-called cavity roundtrip matrix~\cite{DeLasson2014} allows us to determine simultaneously the complex frequency and field distribution of the quasi-normal mode. The iteration is performed using a Newton-Raphson algorithm that requires calculation at three complex wavelengths per iteration. The number of iterations depends on the proximity of the starting guess to the mode wavelength, and in the cases considered here, we have typically needed 8 iterations when the starting guess was within 1 nm of the M1 wavelength.

\subsection{Surface integral equation (SIE) method}

The integral equation method, also known as the Boundary Element Method and the Green's Function Integral Equation Method, solves an integral equation established by enforcing boundary conditions for the electromagnetic field in the frequency domain. Its kernel involves products of Green's functions and some unknown functions. In particular, in the Surface Integral Equation (SIE) method used here, the continuity condition is enforced for the tangential components of the fields at the boundary of an inhomogeneity, in our case the membrane. More specifically, we use the Poggio, Miller, Chew, Harrington, Wu, Tsai (PMCHWT) formulation~\cite{Poggio1973}.


The free-space Green's function in the kernel ensures that the Sommerfeld radiation boundary condition at infinity is inherently satisfied, and as opposed to the previous methods discussed there is no computational domain boundary and thus no need for artificial absorbing boundary conditions. 
{This also means that the SIE approach is not adapted to structures of infinite extent such as a line defect cavity in a finite PhC lattice in an infinite membrane. If such a structure is investigated, the SIE must instead analyze a substructure of finite extent. The error due to the finiteness of the substructure will depend of the penetration of the mode profile to the boundary region of the membrane. For structures with a sufficiently large finite PhC lattice surrounding the cavity, the error may be negligible, while for small surrounding PhC lattices a significant fraction of light may be transmitted to the infinite membrane.}

The integral equation is solved with respect to the equivalent electric and magnetic currents on the surface of the membrane, which can subsequently be used to compute the fields at any point in space, inside or outside the membrane.
To transform the integral equation into a system of linear algebraic equations, the surface of the membrane is divided into a number of higher-order quadrilateral elements to represent the surface of the holes accurately, see Fig.~\ref{Fig:Disc_schemes}(d). The spatial discretization error is controlled by the order $g$ of the Lagrange polynomial interpolation, which ensures smooth and continuous representation of the geometry of each quadrilateral element~\cite{Peterson1998}, and thus stair-casing effects are avoided. The unknown surface currents are expanded in terms of higher-order hierarchical Legendre basis functions providing a higher-order convergence property to the method~\cite{Jørgensen2004}. The polynomial expansion order $p_{\text{max}}$ on each quadrilateral element is dynamically selected according to its electrical size~\cite{Kim2007}. Finally, Galerkin's testing procedure is applied and the resulting linear system is solved by $LU$-decomposition.

Similar to the {aFMM}, a search for the quasi-normal mode is carried out in the complex frequency plane~\cite{Bai2013}. The membrane is excited by a $y$-oriented dipole located at the center of the cavity. The complex frequency $\omega$ is found by searching for a null of $1/|E_y|$, where $E_y$ is the $y$ component of the scattered electric field computed at the point of the source dipole. Alternatively, a real frequency sweep can be used to sample the resonance peak and determine the resonance frequency and $Q$ similarly to sFEM, however in this work we have only used the quasi-normal mode search approach.

Unlike other methods utilizing all three symmetry planes, the SIE utilizes only two, $xz$ and $yz$ planes, simulating one quarter of the structure.

\section{Numerical results: Smaller structure and shorter cavity (L5)} \label{Sec:L5Results}
In this section, we focus on the shorter L5 cavity with in-plane PhC dimensions as given in Table~\ref{Tab:PhCParameters}. For each of the five numerical techniques, we present calculations of the M1 wavelength and $Q$ factor as function of relevant computational parameters. The results are split into two subsections. We first vary the relevant computational domain resolution parameters for all the methods. Subsequently, we study the influence of the size of the computational domain. All presented data is provided in a table format in Appendix A for both cavities.

\subsection{L5: Convergence with numerical resolution} \label{Sec:L5grid}
While all methods employ an increasing number of degrees of freedom to provide a more accurate representation of the geometry and thus a more precise estimate of $\lambda$ and $Q$, the computational parameters used to describe the degrees of freedom vary greatly for the five methods as discussed in Section \ref{Sec:NumericalMethods}. To enable comparison of the results in spite of the different parameters used by the methods, we thus present the results as function of a \textit{common geometrical setup index} $j$. The exact parameters for the geometrical setup indices for each method are presented in Appendix A, but different values of $j$ for FDTD and {aFMM}, for example, correspond to different mesh sizes and Fourier truncations, respectively. Setup 1 contains the lowest number of degrees of freedom and the setup 8 the highest, and convergence for increasing setup index is thus expected. We use the term \textit{resolution} broadly to discuss the performance of the methods when the number of degrees of freedom is varied. {The geometrical setup index was generally chosen by varying a dominating parameter such that the simulation time and the memory requirement display an exponential dependence on the geometrical setup index. An index range including typically a handful of points between a low value and a maximum value allowed by the available computational resources was then chosen. The exception to this procedure was for FDFD and tFEM, where limited computational resources prevented such an exponential stepping, and here a linear variation was chosen instead. This also means that the computational power used for each specific geometrical setup index varies greatly for the different methods. For example, for the highest setup index for the L5 cavity, the discretization step for FDFD was $a/\Delta_x$~=~13, whereas for FDTD it was $a/\Delta_x$~=~438. These discrepancies and their consequences for the comparison of the different methods are discussed further in Section \ref{Sec:Comparison}.}

\begin{figure}[!ht]
	\begin{center}
		\includegraphics[width=13.4cm]{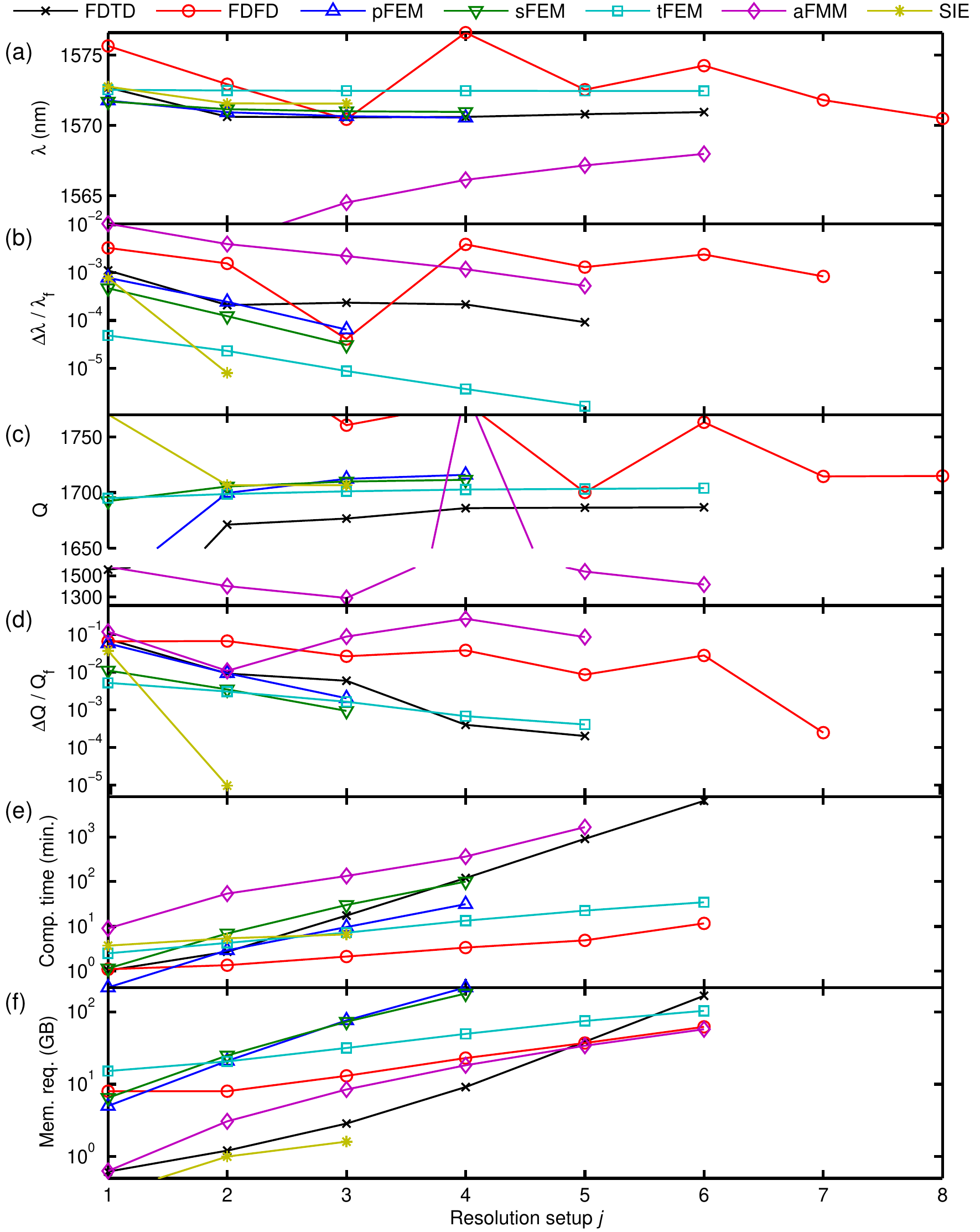}
		\caption{L5: (a) Resonance wavelength $\lambda$, (b) relative wavelength deviation, (c) $Q$ factor, (d) relative $Q$ factor deviation, (e) computation time and (f) memory requirement as function of resolution setup. Notice the disconnected $y$ axis in (c).}
		\label{Fig:L5_grid}
	\end{center}
\end{figure}

In Fig.~\ref{Fig:L5_grid}(a) we study the M1 resonance wavelength $\lambda$ as function of the resolution. The wavelength for the FDTD, pFEM, sFEM, tFEM and SIE methods converges fairly quickly towards a value of $\simeq$ 1571 nm. While the wavelength computed using the FDFD method converges more slowly, the final value is similar to that computed using the other methods. However, the wavelength for the {aFMM} increases uniformly indicating that convergence has not been obtained.

To quantify the degree of convergence, we define the relative deviation $\Delta \lambda _j / \lambda _f$ for the resonance wavelength, where $\Delta \lambda _j = |\lambda_j - \lambda_f|$ is the deviation between the wavelength $\lambda_j$ for the resolution setup $j$ and the "final" wavelength $\lambda_f$ for the setup with the highest resolution. As elaborated in Section \ref{Sec:Discussion}, we stress that the relative deviation should not be considered a precise description of the numerical error in a strict sense, but rather an estimate of the degree of convergence obtained for the specific resolution parameters for each method listed in Appendix A. We then present in Fig.~\ref{Fig:L5_grid}(b) the relative deviation of $\lambda$. The persisting variations of the FDFD and {aFMM} methods lead to the largest relative deviations for these methods with deviations on the order of $10^{-3}$, whereas the deviation for the other methods is generally below $10^{-4}$.

The calculated $Q$ factor and its relative deviation defined similarly as for the wavelength are presented in Figs. \ref{Fig:L5_grid}(c,d). Again the pFEM, sFEM, tFEM and SIE methods converge rapidly to values of $\simeq$ 1710, whereas the FDTD method converges towards a value of 1687. Large variations are again observed for the FDFD technique, which finally converges towards a value similar to that of the pFEM, sFEM, tFEM and SIE methods. The {aFMM} results feature the strongest variations and no clear convergence with a final $Q$ factor of 1417 representing a $\simeq$ 20 \% error compared to the other methods. The pFEM and sFEM methods achieve relative deviations of $\approx 10^{-3}$, whereas for the FDTD and tFEM the deviation is slightly smaller. The SIE approach provides the smallest relative deviation of $10^{-5}$.

Finally, we present the required computation time and the memory requirement in Figs. \ref{Fig:L5_grid}(e,f). Here, the computation time is the total duration from job launch to final $Q$ factor evaluation, and for the iterative methods, the computation time is thus the product of the single iteration calculation time and the number of iterations. 
As discussed earlier, the number of iterations and thus the calculation time will depend on the proximity of the starting guess to the correct value. We observe in Fig. \ref{Fig:L5_grid}(e) that the calculation time generally scales exponentially with the resolution setup index. Typical calculation times are between 10 minutes and 2 hours. However, we notice that the SIE approach achieves a relative deviation for the $Q$ factor below $10^{-5}$ for a simulation time of 5.4 minutes. A similar exponential scaling for the memory requirement can be observed in Fig. \ref{Fig:L5_grid}(f). While the methods typically use between 10 and 100 GB of memory to achieve relative deviation of $10^{-2}$, we observe that the {aFMM} using 64 GB memory is not converged, while the SIE approach achieves a relative deviation below $10^{-5}$ using only 1.6 GB of memory.

\subsection{L5: Dependence on domain size} \label{Sec:L5size}
With the exception of the SIE approach, all the methods use a limited computational domain with PMLs at the boundaries to avoid artificial reflection of light at the domain boundary. For perfectly absorbing PMLs, the $Q$ factor and the resonance wavelength should not depend on the size of the computational domain in the sense that variations in $Q$ and $\lambda$ due to domain size variations should be on the order of the numerical error for the particular resolution setup. In practice, however, the PMLs only suppress the reflection to a certain degree and variations of $Q$ and $\lambda$ as function of the domain size are expected. For this reason, we now study the influence of the size of the computational domain. As for the resolution study, we present our results as function of a \textit{domain size setup index j} with exact geometrical parameters for each setup and method given in Appendix A. Figure \ref{Fig:L5_domainsize} presents results for the FDTD, FDFD, the three FEM methods and the {aFMM}. 

\begin{figure}[b]
	\begin{center}
		\includegraphics[width=13.4cm]{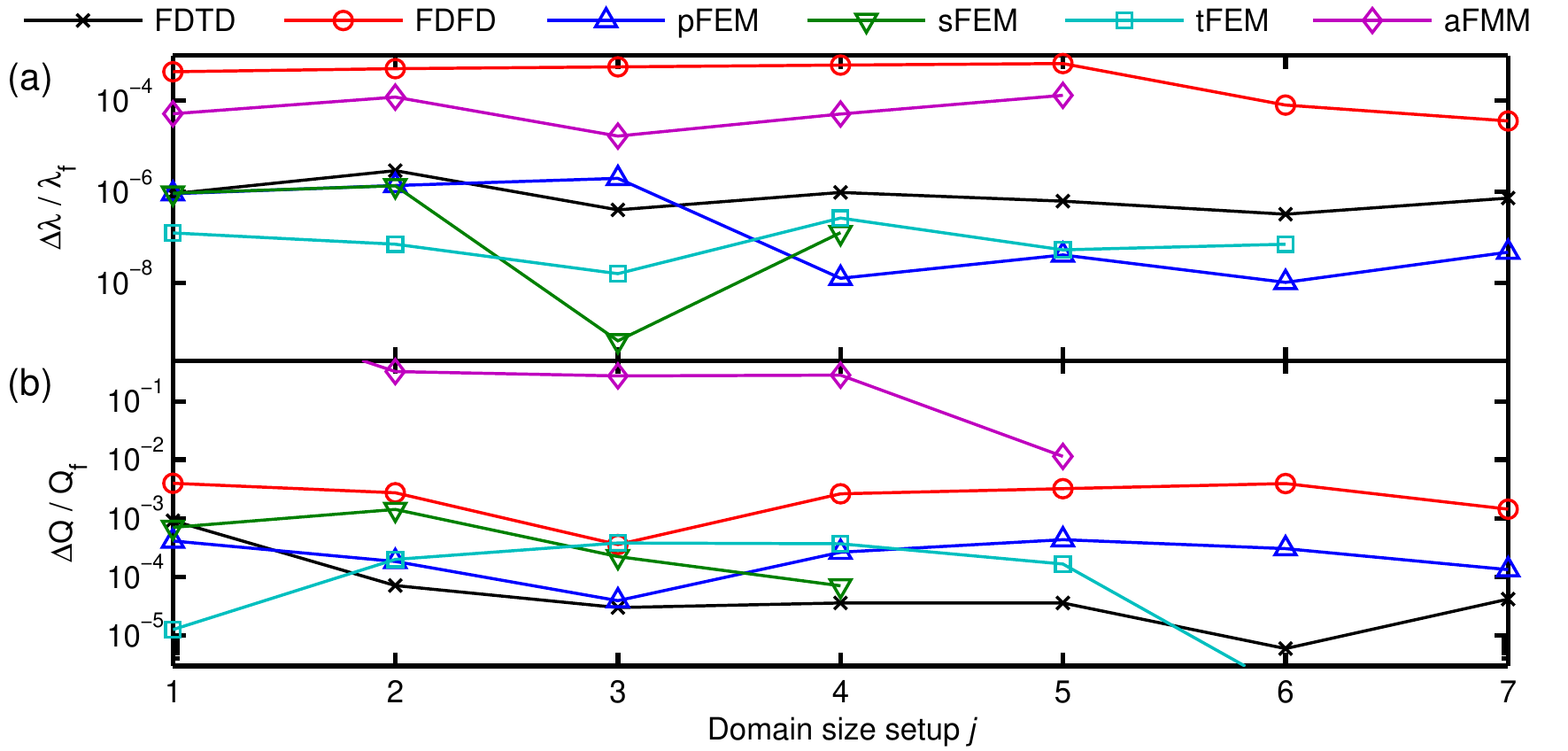}
		\caption{L5: Relative (a) resonance wavelength and (b) $Q$ factor deviations as function of domain setup.}
		\label{Fig:L5_domainsize}
	\end{center}
\end{figure}

The relative deviation in the resonance wavelength is presented in Fig. \ref{Fig:L5_domainsize}(a). We observe that for the FDFD method and the {aFMM}, the PML implementations lead to relative deviations in $\lambda$ around $10^{-4}$, whereas for the other methods the deviation is below $10^{-6}$. As expected, Fig. \ref{Fig:L5_domainsize}(b) shows that the influence of the artificially reflected light is larger on the $Q$ factor with deviations below $10^{-2}$ for the FDFD method, while the deviation is below $10^{-3}$ for the other methods. The $Q$ factor for the {aFMM} features the largest deviation above $\simeq$ $10^{-2}$.

For the L5 cavity, comparison of Figs. \ref{Fig:L5_grid} and \ref{Fig:L5_domainsize} reveals that the error induced by artificial light reflection due to the limited computational domain is smaller than the error due to the finite resolution, and we can conclude that for this geometry the numerical accuracy is generally limited by the resolution rather than by the influence of the boundary conditions.

\section{Numerical results: Larger structure and longer cavity (L9)} \label{Sec:L9Results}
In this section, we consider a longer L9 cavity that is expected to exhibit a higher $Q$ factor than the shorter cavity considered in the previous section~\cite{Okano2010}. Also, as indicated in Table~\ref{Tab:PhCParameters}, the length and width of the PhC membrane are larger. As for the L5 cavity, we study the resonance wavelength and the $Q$ factor first as function of the resolution setup and subsequently as function of the domain setup. 

For the L5 cavity, the convergence was studied for the sFEM as function of $p$ order. However, due to the larger size of the L9 cavity, calculations for $p$ = 4,5 were not feasible. For this reason the convergence for the L9 cavity is only analyzed as function of mesh size $h$ for the sFEM in the following.

\begin{figure}[!t]
	\begin{center}
		\includegraphics[width=13.4cm]{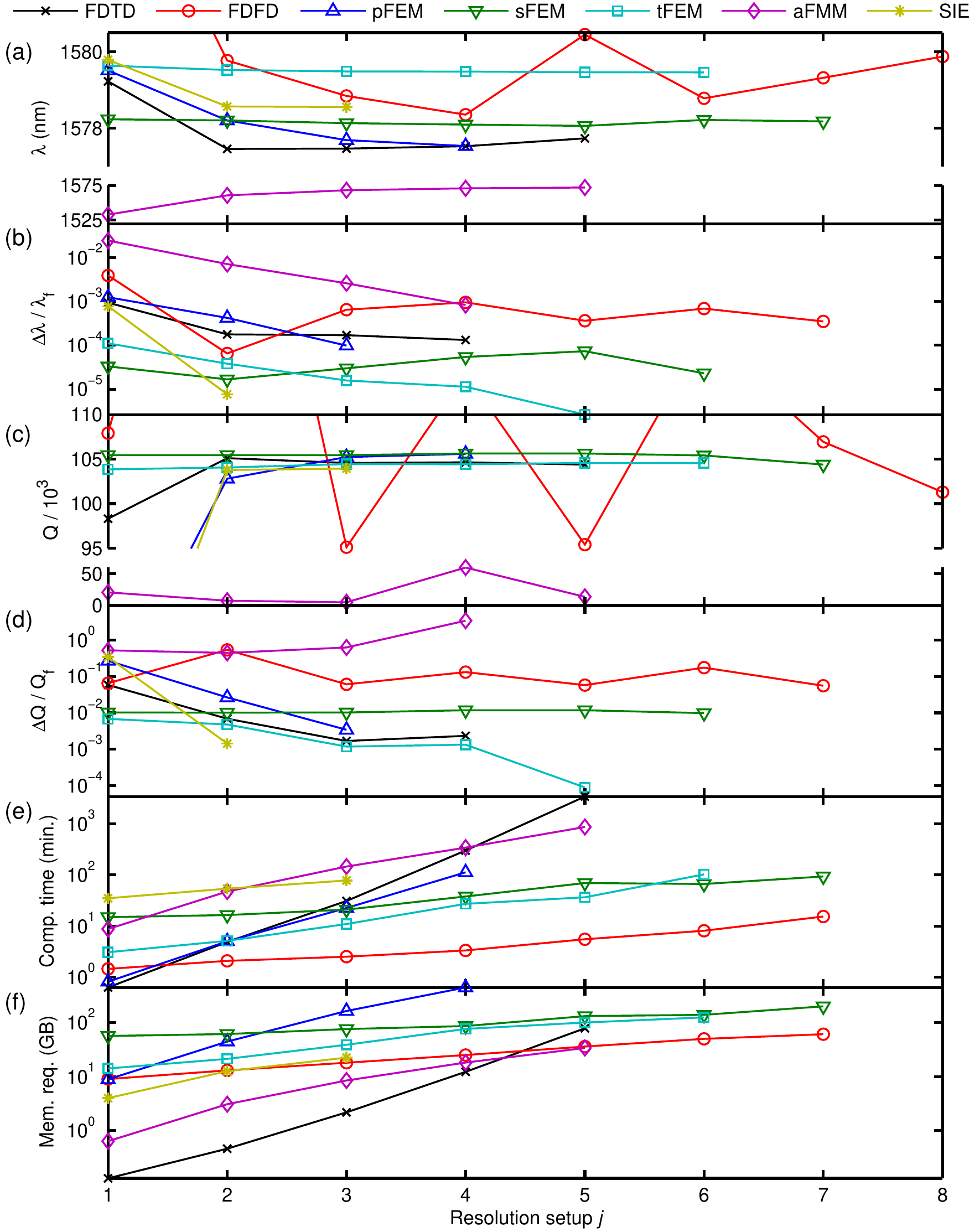}
		\caption{L9: (a) Resonance wavelength $\lambda$, (b) relative wavelength deviation, (c) $Q$ factor, (d) relative $Q$ factor deviation, (e) computation time and (f) memory requirement as function of resolution setup. Notice the disconnected $y$ axis in (a) and (c).}
		\label{Fig:L9_grid}
	\end{center}
\end{figure}

\subsection{L9: Convergence with numerical resolution} \label{Sec:L9grid}

The resonance wavelength $\lambda$ of the L9 cavity is presented in Figs. \ref{Fig:L9_grid}(a,b) as function of resolution setup. The overall behavior resembles that of the L5 cavity, the wavelength for the FDTD, the FEM and the SIE methods converges rapidly with a relative deviations of $\simeq 10^{-5}$ for the tFEM and the SIE methods, whereas the FDFD results display some variations and a relative deviation on the order of $\simeq 10^{-3}$. The FDTD, the FDFD, the FEM and the SIE methods produce similar wavelengths of $\simeq$ 1578 nm. In contrast, the final wavelength computed using {aFMM} is 6 nm smaller, providing a first indication that the {aFMM} has significant difficulty in handling the large L9 cavity.

Results for the $Q$ factor are presented in Figs. \ref{Fig:L9_grid}(c,d). The $Q$ factor from the FDTD, the FEM and the SIE methods converges towards $\simeq$ 104,000. Both the FDFD and the {aFMM} methods display large variations, but whereas the FDFD results appear to converge slowly towards the value predicted by the other methods, the Q factor for the {aFMM} oscillates around an average value of $\approx$ 25.000 and thus deviates by almost a factor of 4 from all other results. The {aFMM} thus appears as the least suitable method for handling the large L9 cavity. The corresponding relative deviations are $\simeq 10^{-1}$ for the FDFD method, while it is below $10^{-2}$ for the pFEM and sFEM and around $10^{-3}$ for the FDTD, tFEM and the SIE methods. 

Whereas the relative deviation generally decreases with the resolution index, we observe the interesting feature that the deviation for the sFEM is stable around $10^{-2}$. The origin of this non-decreasing relative deviation for the sFEM is related to our chosen resonance peak sampling scheme. As discussed in Section \ref{Sec:FEM}, the same number of frequency sampling points was used for both the L5 and the L9 cavities. This number was insufficient to correctly resolve the L9 resonance peak leading to an error due to the sampling scheme of $\simeq$ $10^{-2}$, which dominates over the relative deviation obtained for increasing resolution setup index. This highlights the importance of performing convergence checks for all free parameters used in the simulations, including in this sFEM case the number of iterations $N$ used for the sampling of the resonance.

As before, we present the required computation time and the memory requirement in Figs. \ref{Fig:L9_grid}(e,f). The typical calculation time is around 1-2 hours. We notice that the {aFMM} provides poor convergence even when the simulation time is significantly longer than that of the other methods. The corresponding memory requirement is displayed in Fig. \ref{Fig:L9_grid}(f) and we notice that, not surprisingly, the memory requirement for the L9 cavity is significantly larger than that for the L5 cavity. Again, the SIE method stands out requiring only 22 GB of memory to achieve a relative deviation of $\simeq 10^{-3}$, despite simulating a quarter of the structure, whereas all other methods simulate one eights of it.

\subsection{L9: Dependence on domain size} \label{Sec:L9size}
\begin{figure}[!t]
	\begin{center}
		\includegraphics[width=13.4cm]{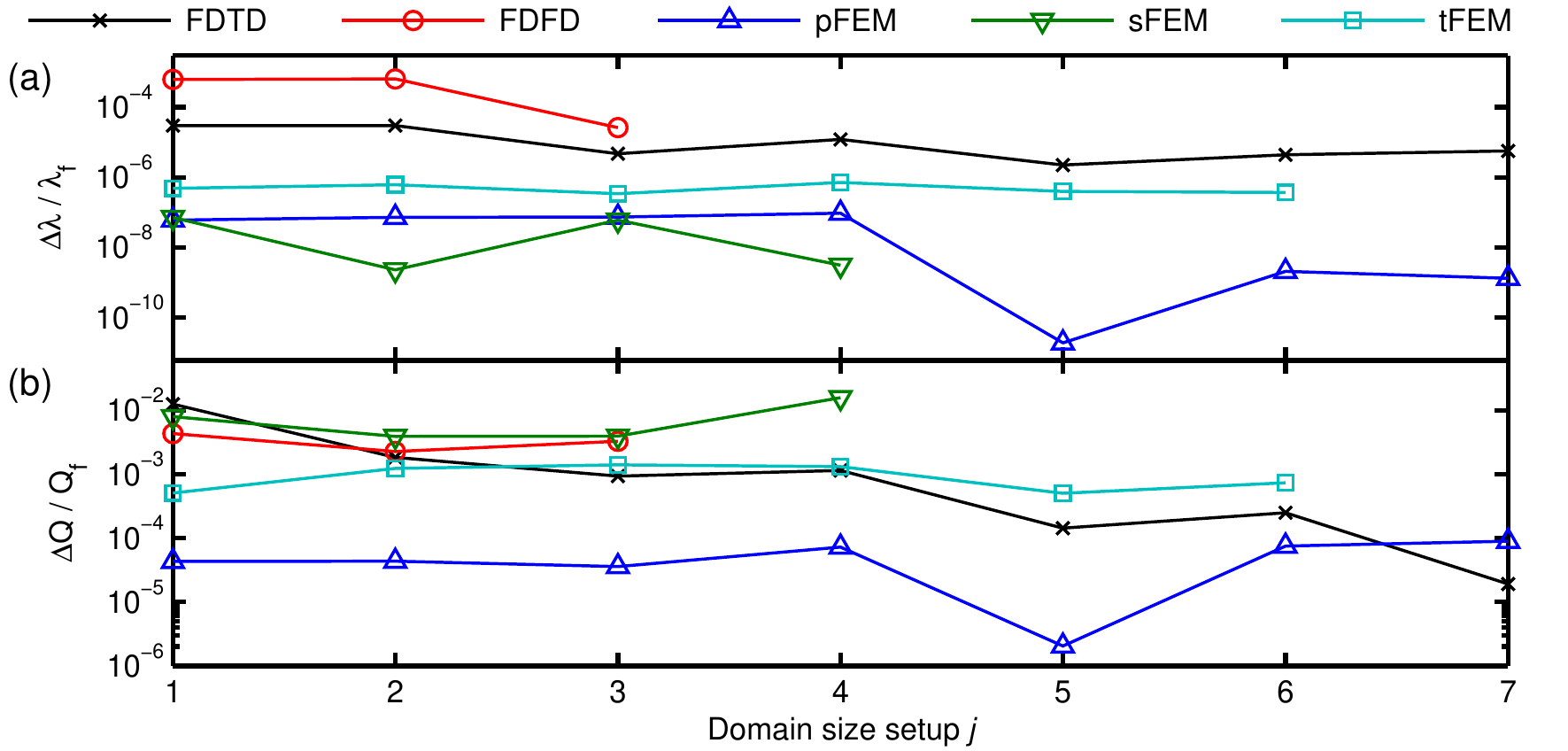}
		\caption{L9: Relative (a) resonance wavelength deviations and (b) $Q$ factor deviations as function of domain setup.}
		\label{Fig:L9_domainsize}
	\end{center}
\end{figure}

As for the L5 cavity, we present convergence results for the L9 cavity as function of the domain size, with geometrical parameters for each setup and method given in Appendix A. The results for the FDTD, FDFD and the three FEM methods are presented in Fig. \ref{Fig:L9_domainsize}. Due to the difficulty of the {aFMM} in obtaining convergence with respect to resolution, convergence studies as function of domain size were not carried out for the {aFMM}.

For the wavelength, the trend is similar to that of the L5 cavity, the FDFD method features the largest deviation of $\simeq 10^{-4}$, whereas deviations around $\simeq 10^{-8}$ are obtained for the pFEM and sFEM methods. 
As before, the relative deviation in the $Q$ factor is larger than that of the wavelength. The FDFD and sFEM methods feature the largest variations with respect to domain size with a deviation of $\simeq 10^{-2}$. On the other hand, deviations of $\simeq 10^{-4}$ are obtained for the FDTD and pFEM methods. 

As for the L5 cavity, the relative deviations for the methods as function of domain size are at least an order of magnitude smaller than that with respect to resolution, indicating that the numerical accuracy for the wavelength and the $Q$ factor are limited mainly by the resolution. The exception is the sFEM method, where the relative error of $\simeq 10^{-2}$ for the domain size study is comparable to that obtained for the resolution study. Here, the error due to an insufficient number of frequencies in the resonance sampling is dominating over those induced by resolution and domain size variations as discussed in Section \ref{Sec:L9grid}.

\begin{table}
	\captionsetup{font=bf}
	\centering 
	\caption{\textbf{Overview of typical computational resources for the L9 calculations.}} 	\label{Tab:Computational_Resources} 
	\begin{threeparttable}
		\begin{tabular}{cccccccc} \hline
			
			& \textbf{FDTD}\tnote{a} & \textbf{FDFD}\tnote{b} & \textbf{pFEM}\tnote{c} & \textbf{sFEM}\tnote{d} & \textbf{tFEM}\tnote{e} & \textbf{{aFMM}}\tnote{f} & \textbf{SIE}\tnote{g} \\  
			\hline 
			Resolution setup \# & L9.R.3 & L9.R.4 & L9.R.2 & L9.R.4 & L9.R.3 & L9.R.3 & L9.R.2 \\
			CPU Cores & 100\tnote{h} & 4\tnote{i} & 10\tnote{j} & 24\tnote{k} & 4\tnote{l} & 8\tnote{h} & 12\tnote{m} \\
			Iterations & 1 & 1 & 1 & 5 & 1 & 8 & 6 \\
			Time/iter. (min) &    &    &    & 7.5 &   & 18 & 9 \\
			Total time (min) & 31 & 3 & 5 & 38 & 11 & 146 & 54 \\		
			Memory (GB) & 2 & 25 & 45 & 86 & 39 & 8 & 12.6 \\ \hline
		\end{tabular}
		\begin{tablenotes}
			\item[a] $a/\Delta x$ = 52, $a/\Delta y$ = 60,  $a/\Delta z$ = 70.1.
			\item[b] $a/\Delta$ = 12.7.	
			\item[c] $p$ = 3, $a/h$ = 1.75. 
			\item[d] $p$ = 3, $a/h$ = 3.8. 
			\item[e] Min. $a/h$ = 1.12, max. $a/h$ = 10.1. 
			\item[f] ($l_y$, $l_z$) = (30, 9), $\lambda_{\text{SG}}=1561$~nm.
			\item[g] $p_\text{max}$=3, $g$=4, $\lambda_{\text{SG}}=1600$~nm.
			{\item[h] Intel Xeon E5-2660v3 @2.6GHz.
				\item[i] Intel Core I7-6700 @3.4GHz. 		
				\item[j] Intel Xeon E5-4650v2 @2.4GHz. 		
				\item[k] Intel Xeon E5-4640 @2.4GHz. 		
				\item[l] Intel Xeon E5-1620v3 @3.5GHz. 
				\item[m] Intel Xeon E5-2650v4 @2.2GHz.}
		\end{tablenotes}
	\end{threeparttable}
\end{table}

\section{Computational resources} \label{Sec:CompRes}
Typical computational resources used to compute the $Q$ factor of the L9 cavity are listed in Table \ref{Tab:Computational_Resources}. The data stem from Table \ref{Tab:L9_NumResData} in Appendix A for the numerical resolution study for the resolution index corresponding to the median of each data set. 

An advantage of the FDTD method is that it can be efficiently parallelized, and the FDTD simulations were performed on a HPC cluster with 100 cores. For the other methods, the potential for exploiting multi-CPU parallelization is smaller and the simulations were performed on single multi-core CPU workstations. The finite elements methods were generally the most memory consuming. However, we remark that only 64 GB and 128 GB of memory were available for the FDFD and the tFEM, respectively, simulations. These methods would certainly benefit from more workstation memory allowing for increased resolution and thus improved convergence. 

\section{Discussion} \label{Sec:Discussion}

When increasing the number of degrees of freedom, we expect convergence of $\lambda$ and $Q$ towards the "true" values for each cavity, which are defined from the geometrical and material parameters alone. However, due to limited computational resources the different methods will not produce these true values exactly, but will instead provide estimates for which the numerical accuracy improves as the number of degrees of freedom is increased. Then, if the numerical error is correctly estimated, the results for the different simulation methods should, in principle, all agree within their respective numerical errors. As we have seen in Sections \ref{Sec:L5Results} and \ref{Sec:L9Results}, the different computational methods provide results with varying relative deviation, and the latter can be considered an estimate of the numerical error for each method. While this definition allows for a quick estimation of the accuracy, it opens up for significant underestimation of the real error as discussed in further detail below.

Ideally, convergence should be studied as function of all numerical parameters simultaneously. For example, for the FDTD method, the parameters $\Delta x$, $\Delta y$, $\Delta z$ (geometrical resolution), $D_x$, $D_y$, and $D_z$ (domain size) should all be varied when estimating the numerical error. This procedure, however, quickly becomes extremely numerically demanding, so instead we have decoupled the resolution and the domain size degrees of freedom. Our general approach is to identify the parameters providing a dominating contribution to the numerical error and estimate the numerical error from variations of these parameters alone. In our study, the error from the resolution parameter variation was generally dominating, and in the following the numerical error estimations are thus based on the resolution index variations.

\begin{figure}
	\begin{center}
		\noindent
		\includegraphics[width=13.4cm]{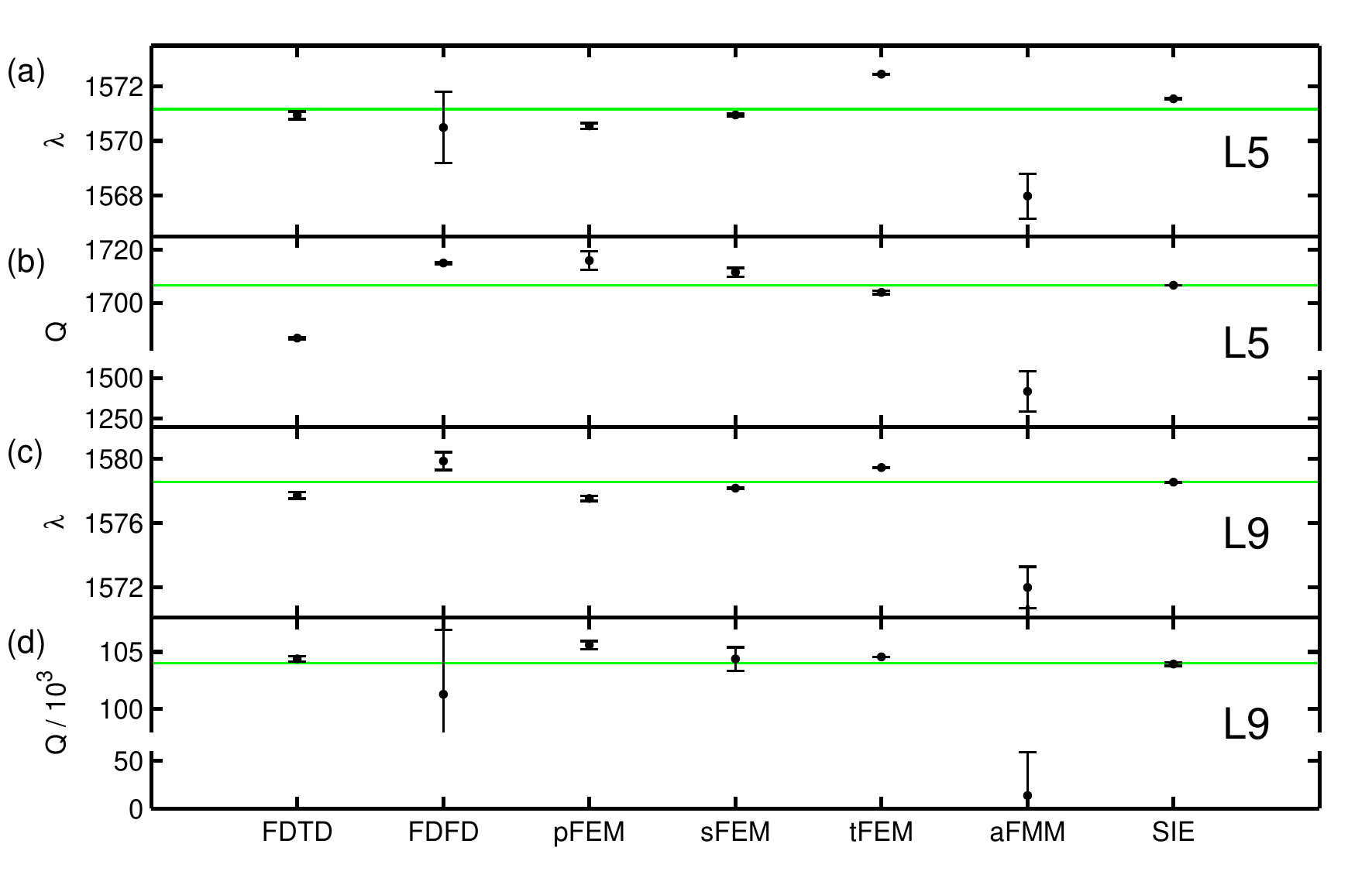}
		\caption{Error bar plot of the resonance wavelength and the $Q$ factor (a,b) for the L5 cavity and (c,d) for the L9 cavity. The green horizontal lines mark average values computed from the FDTD, FDFD, FEM and SIE results. Notice the disconnected $y$ axis in (b) and (d).} \label{Fig:Lx_error}
	\end{center}
\end{figure}

The final parameters $\lambda_f$ and $Q_f$ for the highest resolution indices and the associated error bars are plotted in Fig. \ref{Fig:Lx_error}. Here, the error bars are taken as the relative deviation for the highest index in Figs. \ref{Fig:L5_grid}(b,d) and \ref{Fig:L9_grid}(b,d). As discussed earlier, the {aFMM} results for the $Q$ factor feature the largest deviations from the results of the other methods, with a $\simeq$ 20 \% error for the L5 cavity and a discrepancy compared to all other methods of a factor of four for the L9 cavity. After the {aFMM}, the FDFD results generally display the largest numerical error. We observe that the results for the finite difference techniques, the three FEMs and the SIE approach generally agree fairly well. 

\subsection{Estimation of the numerical error} \label{Sec:NumericalError}

We note that the differences between the results presented in Fig. \ref{Fig:Lx_error} in many cases are significantly larger than the error bars. From this, we conclude that a quantitative estimate of the numerical error from the relative deviation leads to an underestimation of the real numerical error of the methods. Several factors contribute to this underestimation as explained in the following.

\begin{figure}[!htb]
	\begin{center}
		\noindent
		\includegraphics[width=13.4cm]{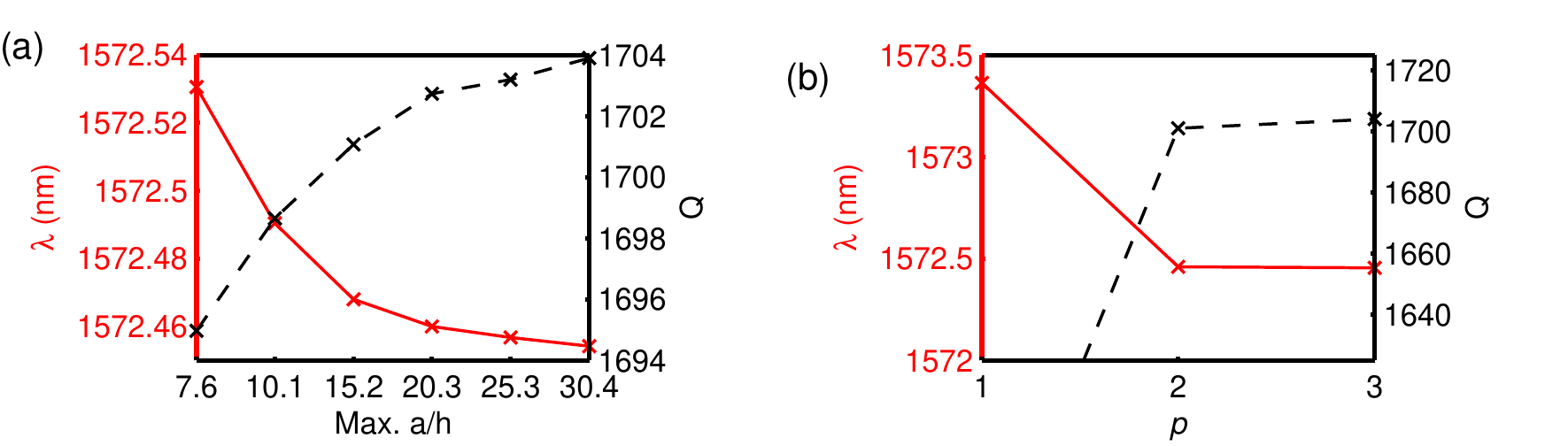}
		\caption{L5, tFEM: $Q$ and $\lambda$ (a) as function of  mesh size for $p=2$ and (b) as function of polynomial order $p$ for resolution setup 4. The (a) data is identical to those in Fig. \ref{Fig:L5_grid}.}
		\label{Fig:COMSOL_p_order}
	\end{center}
\end{figure}

A first factor is the lack of variation of a parameter with dominating influence on the numerical error in the convergence study. For example, for the tFEM calculations the polynomial order $p=2$ was kept fixed and the mesh size $h$ was varied. However, the $p$ order in FEM simulations has a significant impact on the result. In Fig. \ref{Fig:COMSOL_p_order}, the influence of the mesh size and the $p$ order on $Q$ and $\lambda$ are presented. Whereas $Q$ varies by 9 in the mesh order study in Fig. \ref{Fig:COMSOL_p_order}(a), the variation is $\simeq$ 170 in the $p$-order study in Fig. \ref{Fig:COMSOL_p_order}(b). A more accurate estimation of the error could thus be performed by performing a $p$-order study or by at least setting $p = 3$ in the mesh size variation study. However, since only 128 GB of memory was available for the tFEM simulations, this was not possible. Furthermore, COMSOL does not support $p>3$. 

Another reason for the error underestimation is that the magnitude of the relative deviations and thus the error bars in Fig. \ref{Fig:Lx_error} depend on the variation of the resolution setup parameters. While in the case of the FDTD method, the resolution index was varied exponentially, the variation was linear for the FDFD, tFEM, sFEM (L9) and {aFMM}. Since the relative deviation can be reduced arbitrarily by decreasing the incremental resolution step, another description of the numerical error for linear variation of the resolution indices is clearly needed. In the case of oscillatory variations as function of the resolution index, a more accurate definition can be the maximum deviation from an average value during a full oscillatory cycle. However, this procedure requires more data points than we had available. 

Furthermore, the actual structures analyzed by the different methods are not always identical. In the case of sFEM calculations for the L5 structure, a $p$-order study was performed for the mesh illustrated in Fig. \ref{Fig:Disc_schemes}(b). While a high numerical accuracy may be obtained by increasing $p$, the convergence study is performed not for the ideal structure but for the representation of the circular holes using the prism mesh. 
{Similarly, the geometry analyzed by the aFMM was not the true structure but the staircase approximation shown Fig. \ref{Fig:Disc_schemes}(c).}
The mesh resolution {for the FEM and the number of steps in the staircase approximation for the aFMM were} chosen such that the error due to inaccurate representation of the holes was estimated to be negligible, but deviations in the results due to differences in the geometrical representations of the different methods may occur and are not always correctly captured in the estimation of the error.

\subsection{Comparison of the methods} \label{Sec:Comparison}

The differences in the $Q$ factors and the resonance wavelengths displayed in Fig. \ref{Fig:Lx_error} indicate that all methods have produced results deviating from the true ones to a larger or lesser degree. While differences might be expected due to the use of different methods, our benchmark study of the 3 FEMs indicate that deviations occur even for different implementations of the same method. In addition, a surprising - and disturbing - realization is that the numerical error in many cases is vastly underestimated. This lack of confidence in the numerical results is rarely discussed in the literature \cite{Granet2006}, even though it should be a major concern when studying convergence of numerical simulations. With this issue in mind, one may now ask the question: Which methods are suitable or not for computing resonance wavelengths and $Q$ factors in large nanophotonic cavities? 

In our study, independent simulations were performed by the different groups involved in this work using the computational hardware available to each group.
{The difference in hardware resources has meant that the feasible parameter ranges for the studies of the influence of the numerical resolution and of the domain size were different for each method.}
In principle, this procedure minimizes the risk of systematic error, and the quantitative agreement may be regarded as more convincing as it has been reached independently. However, it does also mean that in the comparison we should keep in mind that different computational resources were available in the benchmarking of each method. Thus, while the change in computation time with the resolution setup can thus be compared for each method individually, when comparing the computation time for the various methods one should keep in mind that different CPUs with varying clock frequencies were used. Similarly, the convergence of the FDFD and tFEM simulations was clearly limited by the 64 GB and 128 GB of memory in the workstations used. As discussed earlier, this limitation manifests itself, for example, when comparing the grid resolution used in the FDFD and FDTD methods: While the highest resolution was $a/\Delta x$ = 22.2 for the FDFD simulations, it was an order of magnitude larger, $a/\Delta x$ = 438, for the FDTD calculations for the L5 cavity. Also, even though the grid resolutions for the FDFD and the tFEM appear similar, for the same side length the volume of the cubic cell used in FDFD is twice as large as that of the tetrahedral element used in tFEM. Finally, four times as much memory was available for the pFEM simulations as compared to tFEM.

Keeping these differences in available resources in mind, our study clearly indicates that the least suitable method is the {aFMM}. The {aFMM} is based on plane-wave expansion of the permittivity profile as well as of the electromagnetic fields, and the poor convergence is likely due to the difficulty of accurately describing the large electric field discontinuities occurring at the InP-air interfaces using this series expansion~\cite{Sozuer1992}. On the other hand, these field discontinuities are handled without any problem by the finite difference and finite elements methods which in this respect perform well, but at the cost of significant demands to memory, CPU cores and calculation time. 
{Finally, the SIE approach}
provides excellent convergence using the smallest memory requirement of all the methods. 
The impressive performance of the SIE approach for these PhC line defect cavities is explained by two factors: It is based on an accurate description of the structure with a discretization of the surface, cf.\ Fig.~\ref{Fig:Disc_schemes}(d), of the perforated membrane with higher-order quadrilateral elements and additionally, the SIE approach features an inherent open geometry boundary condition. 
These features make the SIE approach very suitable for the studied line defect cavities
{of finite extent, while structures of infinite extent cannot be handled with the SIE used here.}

We have included three implementations, pFEM, sFEM and tFEM, of the finite elements method, and again a fair comparison should take into account the different computational resources available for the calculations. For the tFEM, a very low relative deviation of $10^{-4}$ for the $Q$ factor of the L9 cavity was obtained, but as discussed earlier this low number is most likely due to the incremental change in resolution between setups L9.R.5 and L9.R.6. Similar incremental modification of the resolution index was used for the L9 sFEM simulations, however here the larger relative deviation $10^{-2}$ was most likely caused by an insufficient number $N$ of iterations for the cavity resonance. Increasing $N$ will lead to improved accuracy but at the cost of increased simulation time, which is already $\simeq$ 3 times larger (L5) than for the pFEM. We thus conclude that the eigenvalue technique appears favorable compared to the scattering approach for the FEM. While a direct comparison of the performance of the pFEM and tFEM is hampered by the different amounts of memory available, we have demonstrated the importance of the $p$-order convergence study and here the lack of support for $p>3$ in COMSOL is clearly a weakness.

Finally we note that the choice of the best numerical method for a specific simulation task depends on the required numerical accuracy, and on the type of simulation task. The specific parameter setting plays an important role as, e.g., in general the computation costs of the different methods scale differently with the size of the computational domain. Furthermore, the functionality of the device is important: Resonators are the most sensitivities structures imaginable, and this explains the rather large deviations between the different methods observed in this study.

\section{Conclusion and outlook} \label{Sec:Conclusion}
We have used five numerical simulation methods and three implementations of the finite elements method to compute $Q$ factors and resonance wavelengths of photonic crystal L5 and L9 line defect cavities. We have provided detailed studies of the influence of the resolution parameters as well as of the domain size, and we have compared the convergence of the methods. While we conclude that the {worst performance is obtained using the {aFMM}}, more surprising discoveries were the disagreement between the various methods and the difficulty of correctly evaluating the numerical error. 

Based on these findings, we hope that the challenge of establishing confidence in numerical results will gain more attention in the photonic community and literature, and we encourage further discussion of how to correctly evaluate numerical accuracy and how to make fair performance comparisons of different simulation methods when analyzing highly challenging nanophotonic components.

\section*{Appendix A: Detailed simulation data} 

In the following tables, we employ the coding L$n$.$X$.$j$ where $n$ = 5,9 and $X$ = R,D for the Resolution/Domain setup index.

Analysis of stability and speed of the iterative algorithms used to determine complex quasi-normal mode frequencies using the {aFMM} and SIE methods is outside the scope of the present paper. However, for completeness we include the iteration start guesses that have led to the computed quasi-normal mode frequencies. 

\begin{table}[H]
	\centering 
	\captionsetup{font=bf}
	\caption{\textbf{Data for the numerical resolution study for the L5 cavity.}}	
	\begin{threeparttable}
		\begin{tabular}{cccccccccc} \hline
			Setup & \multicolumn{3}{c}{\multirow{2}{*}{Resolution}} & \multirow{2}{*}{$\lambda$ (nm)} & \multirow{2}{*}{$Q$} & Time & Mem. & Domain \\
			\# & & & & & & (min.) & (GB) & size \\ \hline
			\textbf{FDTD} & $a/\Delta x$ & $a/\Delta y$ & $a/\Delta z$ \\
			L5.R.1 & 13 & 15 & 17.5 & 1572.69 & 1558.82 & 1.033 & 0.618  \\
			L5.R.2 & 26 & 30 & 35 & 1570.61 & 1671.18 & 2.717 & 1.21 & $D_x/a$ = 40 \\
			L5.R.3 & 52 & 60 & 70.1 & 1570.57 & 1676.68 & 17.42 & 2.85  & $D_y/a$ = 41.6 \\
			L5.R.4 & 104 & 120 & 140 & 1570.6 & 1686.07 & 120.4 & 9.1 & $D_z/a$ = 25 \\
			L5.R.5 & 208 & 240 & 280 & 1570.8 & 1686.41 & 902.5 & 38.5 \\
			L5.R.6 & 438 & 506 & 438 & 1570.94 & 1686.75 & 6390 & 167 \\
			\hline
			\textbf{FDFD} & $a / \Delta x$ & $a / \Delta y$ & $a / \Delta z$ \\
			L5.R.1 & 10.6 & 10.6 & 10.6 & 1575.65 & 1829.12 & 1.108 & 8 \\
			L5.R.2 & 11.5 & 11.5 & 11.5 & 1572.93 & 1831.63 & 1.358 & 8 \\
			L5.R.3 & 12.5 & 12.5 & 12.5 & 1570.43 & 1760.47 & 2.125 & 13 & $D_x/a$ = 6.5 \\
			L5.R.4 & 13.7 & 13.7 & 13.7 & 1576.61 & 1780.52 & 3.367 & 23 & $D_y/a$ = 3.6 \\
			L5.R.5 & 15.2 & 15.2 & 15.2 & 1572.54 & 1700.05 & 4.86 & 37  & $D_z/a$ = 1.3 \\
			L5.R.6 & 16.9 & 16.9 & 16.9 & 1574.26 & 1763.1 & 11.68 & 62 \\
			L5.R.7 & 19.2 & 19.2 & 19.2 & 1571.8 & 1714.49 & N/A & N/A \\
			L5.R.8 & 22.2 & 22.2 & 22.2 & 1570.49 & 1714.91 & N/A & N/A \\
			\hline
			\textbf{pFEM} & $p$ &  $a/h$ \\
			L5.R.1 & 2 & 2.9 &  & 1571.77 & 1615.22 & 0.4333 & 5 & $D_x/a$ = 0 \\
			L5.R.2 & 3 & 2.9 &  & 1570.93 & 1699.61 & 2.917 & 20.9 & $D_y/a$ = 0  \\
			L5.R.3 & 4 & 2.9 &  & 1570.65 & 1712.35 & 9.667 & 76 & $D_z/a$ = 0  \\
			L5.R.4 & 5 & 2.9 &  & 1570.55 & 1715.87 & 31.43 & 218 \\
			\hline
			\textbf{sFEM} & $p$ &  $a/h$ \\
			L5.R.1 & 2 & 3.4 &  & 1571.7 & 1692.32 & 1.152 & 6.47 & $D_x/a$ = 0  \\
			L5.R.2 & 3 & 3.4 &  & 1571.15 & 1705.43 & 6.928 & 24.9 & $D_y/a$ = 0  \\
			L5.R.3 & 4 & 3.4 &  & 1571 & 1709.85 & 30.18 & 73 & $D_z/a$ = 0  \\
			L5.R.4 & 5 & 3.4 &  & 1570.96 & 1711.46 & 100.4 & 179 \\
			\hline
			\textbf{tFEM} & $p$ & Min. $a/h$ & Max. $a/h$ \\
			L5.R.1 & 2 & 1.12 & 7.6 & 1572.53 & 1694.97 & 2.5 & 15.2 \\
			L5.R.2 & 2 & 1.12 & 10.1 & 1572.49 & 1698.66 & 4.267 & 20.7 & $D_x/a$ = 4 \\
			L5.R.3 & 2 & 1.12 & 15.2 & 1572.47 & 1701.08 & 7.217 & 31.6 & $D_y/a$ = 4 \\
			L5.R.4 & 2 & 1.12 & 20.3 & 1572.46 & 1702.74 & 13.38 & 49.6 & $D_z/a$ = 4 \\
			L5.R.5 & 2 & 1.12 & 25.3 & 1572.46 & 1703.2 & 22.52 & 75.1 \\
			L5.R.6 & 2 & 1.12 & 30.4 & 1572.45 & 1703.9 & 34.6 & 103 \\
			\hline
			\textbf{{aFMM}} & $l_y$ & $l_z$ & $\lambda_{\text{SG}}$ (nm) \\
			L5.R.1 & 10 & 7 & 1548 & 1551.41 & 1586.2 & 9.017 & 0.63 \\
			L5.R.2 & 20 & 8 & 1558 & 1561.77 & 1401.65 & 54.03 & 3.06 \\
			L5.R.3 & 30 & 9 & 1561 & 1564.49 & 1290.34 & 133.8 & 8.46 & $D_y/a$ = 0.86 \\
			L5.R.4 & 40 & 10 & 1563 & 1566.12 & 1791.37 & 361.9 & 18.2 & $D_z/a$ = 0.86 \\
			L5.R.5 & 50 & 11 & 1563 & 1567.15 & 1539.87 & 1655 & 33.9 \\
			L5.R.6 & 60 & 12 & N/A & 1567.97 & 1417.08 &  & 57.5 \\
			\hline
			\textbf{SIE} & $p_\text{max}$ & $g$ & $\lambda_{\text{SG}}$ (nm) \\
			L5.R.1 & 2 & 4 & 1600 & 1572.76 & 1770.25 & 3.7 & 0.3 \\
			L5.R.2 & 3 & 4 & 1600 & 1571.56 & 1706.56 & 5.4 & 1 \\
			L5.R.3 & 4 & 4 & 1600 & 1571.55 & 1706.58 & 6.6 & 1.6 \\
			\hline
		\end{tabular}
	\end{threeparttable}	
	\label{Tab:L5_NumResData} 
\end{table}

\begin{table}[H]
	\centering 
	\captionsetup{font=bf}	
	\caption{\textbf{Data for the domain size study for the L5 cavity.}}	
	\begin{threeparttable}
		\begin{tabular}{cccccccc} \hline
			Setup \# & $D_x/a$ & $D_y/a$ & $D_z/a$ & $\lambda$ (nm) & $Q$ & \multicolumn{2}{c}{Resolution}  \\ \hline
			\textbf{FDTD}  \\
			L5.D.1 & 5 & 6.9282 & 7.5 & 1570.61 & 1687.55 \\
			L5.D.2 & 10 & 13.8564 & 10 & 1570.61 & 1685.9 \\
			L5.D.3 & 20 & 27.7128 & 15 & 1570.6 & 1686.07 & \multicolumn{2}{c}{$a/ \Delta x$ = 104}  \\
			L5.D.4 & 30 & 41.5692 & 20 & 1570.61 & 1685.96 & \multicolumn{2}{c}{$a/ \Delta y$ = 120}  \\
			L5.D.5 & 40 & 55.4256 & 25 & 1570.6 & 1686.08 & \multicolumn{2}{c}{$a/ \Delta z$ = 140}  \\
			L5.D.6 & 50 & 69.282 & 30 & 1570.61 & 1686.01 \\
			L5.D.7 & 60 & 83.1384 & 35 & 1570.6 & 1686.09 \\
			L5.D.8 & 70 & 96.9948 & 40 & 1570.6 & 1686.02 \\
			\hline
			\textbf{FDFD}  \\
			L5.D.1 & 0.178 & 0.178 & 0.553 & 1571.57 & 1713.2 \\
			L5.D.2 & 0.231 & 0.231 & 0.606 & 1571.69 & 1715.3 \\
			L5.D.3 & 0.284 & 0.284 & 0.659 & 1571.77 & 1719.38  & \multicolumn{2}{c}{$a/ \Delta x$ = 18.9} \\
			L5.D.4 & 0.337 & 0.337 & 0.712 & 1571.84 & 1715.48  & \multicolumn{2}{c}{$a/ \Delta y$ = 18.9}  \\
			L5.D.5 & 0.39 & 0.39 & 0.765 & 1571.92 & 1714.48  & \multicolumn{2}{c}{$a/ \Delta z$ = 18.9} \\
			L5.D.6 & 0.443 & 0.443 & 0.818 & 1570.76 & 1713.21 \\
			L5.D.7 & 0.496 & 0.496 & 0.871 & 1570.83 & 1722.45 \\
			L5.D.8 & 0.549 & 0.549 & 0.924 & 1570.89 & 1719.99 \\
			\hline
			\textbf{pFEM}  \\
			L5.D.1 & 0.11 & 0.11 & 0.11 & 1570.93 & 1705.33 \\
			L5.D.2 & 0.23 & 0.23 & 0.23 & 1570.94 & 1704.94 \\
			L5.D.3 & 0.46 & 0.46 & 0.46 & 1570.94 & 1704.56 \\
			L5.D.4 & 0.68 & 0.68 & 0.68 & 1570.93 & 1704.18 & \multicolumn{2}{c}{$p$ = 3} \\
			L5.D.5 & 0.91 & 0.91 & 0.91 & 1570.93 & 1703.89 & \multicolumn{2}{c}{$a/h$ = 2.9} \\
			L5.D.6 & 1.14 & 1.14 & 1.14 & 1570.93 & 1704.11 \\
			L5.D.7 & 1.71 & 1.71 & 1.71 & 1570.93 & 1704.85 \\
			L5.D.8 & 2.28 & 2.28 & 2.28 & 1570.93 & 1704.63 \\
			\hline
			\textbf{sFEM}  \\
			L5.D.1 & 0 & 0 & 0 & 1571.15 & 1705.43 \\
			L5.D.2 & 0.02 & 0.02 & 0.02 & 1571.15 & 1706.67 & \multicolumn{2}{c}{$p$ = 3} \\
			L5.D.3 & 0.23 & 0.23 & 0.23 & 1571.15 & 1704.62 & \multicolumn{2}{c}{$a/h$ = 3.4} \\
			L5.D.4 & 0.46 & 0.46 & 0.46 & 1571.15 & 1704.13 \\
			L5.D.5 & 1.14 & 1.14 & 1.14 & 1571.15 & 1704.25 \\
			\hline
			\textbf{tFEM}  \\
			L5.D.1 & 4 & 4 & 4 & 1572.46 & 1702.72 & \multicolumn{2}{c}{$p$ = 2} \\
			L5.D.2 & 4 & 4 & 6 & 1572.46 & 1702.4 \\
			L5.D.3 & 4 & 4 & 8 & 1572.46 & 1702.1 & \multicolumn{2}{c}{Min. $a/h$}  \\
			L5.D.4 & 4 & 4 & 10 & 1572.46 & 1702.11 & \multicolumn{2}{c}{= 1.12} \\
			L5.D.5 & 4 & 4 & 12 & 1572.46 & 1702.46 \\
			L5.D.6 & 4 & 4 & 14 & 1572.46 & 1702.74 & \multicolumn{2}{c}{Max. $a/h$}  \\
			L5.D.7 & 4 & 4 & 16 & 1572.46 & 1702.74 & \multicolumn{2}{c}{= 20.3} \\
			\hline 
			\textbf{{aFMM}} & & & & & & $l_y$ & $l_z$ \\
			L5.D.1 & $\infty$ & 0.51 & 0.51 & 1565.96 & 10297.7 & 37 & 8 \\
			L5.D.2 & $\infty$ & 0.63 & 0.63 & 1566.23 & 951.075 & 38 & 9 \\
			L5.D.3 & $\infty$ & 0.74 & 0.74 & 1566.01 & 1020.36 & 39 & 9 \\
			L5.D.4 & $\infty$ & 0.86 & 0.86 & 1566.12 & 1791.37 & 40 & 10 \\
			L5.D.5 & $\infty$ & 0.97 & 0.97 & 1566.25 & 1384.39 & 41 & 11 \\
			L5.D.6 & $\infty$ & 1.08 & 1.08 & 1566.04 & 1400.45 & 42 & 11 \\
			\hline
		\end{tabular}
	\end{threeparttable}	
	\label{Tab:L5_DomainSizeData} 
\end{table}

\begin{table}[H]
	\centering
	\captionsetup{font=bf}	
	\caption{\textbf{Data for the numerical resolution study for the L9 cavity.}}	
	\begin{threeparttable}
		\begin{tabular}{cccccccccc} \hline
			Setup & \multicolumn{3}{c}{\multirow{2}{*}{Resolution}} & \multirow{2}{*}{$\lambda$ (nm)} & \multirow{2}{*}{$Q$} & Time & Mem. & Domain \\
			\# & & & & & & (min.) & (GB) & size \\ \hline
			\textbf{FDTD} & $a/\Delta x$ & $a/\Delta y$ & $a/\Delta z$ \\
			L9.R.1 & 13 & 15 & 17.5 & 1579.22 & 98311.7 & 0.6333 & 0.131 \\
			L9.R.2 & 26 & 30 & 35 & 1577.45 & 105106 & 4.883 & 0.46 & $D_x/a$ = 5 \\
			L9.R.3 & 52 & 60 & 70.1 & 1577.46 & 104565 & 31.02 & 2.17 & $D_y/a$ = 2.60 \\
			L9.R.4 & 104 & 120 & 140.2 & 1577.52 & 104632 & 300.4 & 12.2 & $D_z/a$ = 3 \\
			L9.R.5 & 208 & 240 & 280.3 & 1577.73 & 104389 & 3422 & 78.3 \\
			\hline
			\textbf{FDFD} & $a / \Delta x$ & $a / \Delta y$ & $a / \Delta z$ \\
			L9.R.1 & 10.3 & 10.3 & 10.3 & 1586.1 & 107906 & 1.455 & 9 \\
			L9.R.2 & 11 & 11 & 11 & 1579.77 & 155805 & 2.088 & 13 \\
			L9.R.3 & 11.8 & 11.8 & 11.8 & 1578.84 & 95124 & 2.522 & 18 & $D_x/a$ = 11.2 \\
			L9.R.4 & 12.7 & 12.7 & 12.7 & 1578.35 & 114578 & 3.327 & 25 & $D_y/a$ = 5.8 \\
			L9.R.5 & 13.7 & 13.7 & 13.7 & 1580.45 & 95412.8 & 5.548 & 36 & $D_z/a$ = 1.3 \\
			L9.R.6 & 14.9 & 14.9 & 14.9 & 1578.78 & 119089 & 8.082 & 50 \\
			L9.R.7 & 16.4 & 16.4 & 16.4 & 1579.31 & 106936 & 15.39 & 61 \\
			L9.R.8 & 18.2 & 18.2 & 18.2 & 1579.87 & 101289 & N/A & N/A \\
			\hline
			\textbf{pFEM} & $p$ &  $a/h$ \\
			L9.R.1 & 2 & 1.75 &  & 1579.51 & 77096.1 & 0.8167 & 8.9 & $D_x/a$ = 0 \\
			L9.R.2 & 3 & 1.75 &  & 1578.2 & 102805 & 5.017 & 44.6 & $D_y/a$ = 0 \\
			L9.R.3 & 4 & 1.75 &  & 1577.69 & 105233 & 22.63 & 164 & $D_z/a$ = 0 \\
			L9.R.4 & 5 & 1.75 &  & 1577.53 & 105595 & 112.7 & 449 \\
			\hline
			\textbf{sFEM} & $p$ &  $a/h$ \\
			L9.R.1 & 3 & 2.2 &  & 1578.23 & 105454 & 15.01 & 56.3 \\
			L9.R.2 & 3 & 2.7 &  & 1578.2 & 105440 & 16.47 & 61.4 \\
			L9.R.3 & 3 & 3.4 &  & 1578.13 & 105449 & 20.97 & 75.7 & $D_x/a$ = 0 \\
			L9.R.4 & 3 & 3.8 &  & 1578.09 & 105621 & 37.86 & 86.6 & $D_y/a$ = 0 \\
			L9.R.5 & 3 & 4.4 &  & 1578.06 & 105617 & 69.42 & 132 & $D_z/a$ = 0 \\
			L9.R.6 & 3 & 4.9 &  & 1578.21 & 105416 & 66.49 & 139 \\
			L9.R.7 & 3 & 5.5 &  & 1578.18 & 104387 & 93.33 & 200 \\
			\hline
			\textbf{tFEM} & $p$ & Min. $a/h$ & Max. $a/h$ \\
			L9.R.1 & 2 & 1.12 & 5.1 & 1579.63 & 103848 & 3.083 & 14.2 \\
			L9.R.2 & 2 & 1.12 & 7.6 & 1579.52 & 104064 & 5.133 & 21.2 & $D_x/a$ = 4 \\
			L9.R.3 & 2 & 1.12 & 10.1 & 1579.48 & 104438 & 10.95 & 38.9 & $D_y/a$ = 4 \\
			L9.R.4 & 2 & 1.12 & 12.7 & 1579.48 & 104423 & 27.25 & 76.5 & $D_z/a$ = 4 \\
			L9.R.5 & 2 & 1.12 & 15.2 & 1579.46 & 104571 & 36.67 & 101 \\
			L9.R.6 & 2 & 1.12 & 17.7 & 1579.46 & 104562 & 102.8 & 125 \\
			\hline
			\textbf{{aFMM}} & $l_y$ & $l_z$ & $\lambda_\text{SG}$ (nm) \\
			L9.R.1 & 10 & 7 & 1532 & 1533.12 & 20567.1 & 8.75 & 0.638 \\
			L9.R.2 & 20 & 8 & 1565 & 1560.68 & 7516.90 & 47.45 & 3.07 & $D_y/a$ = 0.86  \\
			L9.R.3 & 30 & 9 & 1561 & 1567.89 & 4991.99 & 145.7 & 8.46 & $D_z/a$ = 0.86  \\
			L9.R.4 & 40 & 10 & 1563 & 1570.67 & 59499.3 & 341 & 18.2 \\
			L9.R.5 & 50 & 11 & 1563 & 1571.98 & 13507.2 & 869 & 34.0 \\
			\hline
			\textbf{SIE} & $p_\text{max}$ & $g$ & $\lambda_\text{SG}$ (nm) \\
			L9.R.1 & 2 & 4 & 1600 & 1579.79 & 68505.8 & 35 & 4 \\
			L9.R.2 & 3 & 4 & 1600 & 1578.57 & 103782 & 54 & 12.6 \\
			L9.R.3 & 4 & 4 & 1600 & 1578.55 & 103931 & 78 & 22.4 \\
			\hline
		\end{tabular}
	\end{threeparttable}	
	\label{Tab:L9_NumResData} 
\end{table}

\begin{table}[H]
	\centering
	\captionsetup{font=bf}	
	\caption{\textbf{Data for the domain size study for the L9 cavity.}}	
	\begin{threeparttable}
		\begin{tabular}{ccccccc} \hline
			Setup \# & $D_x/a$ & $D_y/a$ & $D_z/a$ & $\lambda$ (nm) & $Q$ & Resolution  \\ \hline
			\textbf{FDTD}  \\
			L9.D.1 & 1 & 1.1547 & 1 & 1577.58 & 103315 \\
			L9.D.2 & 2 & 1.7321 & 1.5 & 1577.58 & 104807 \\
			L9.D.3 & 3 & 2.3094 & 2 & 1577.54 & 104713 & $a/ \Delta x$ = 104 \\
			L9.D.4 & 4 & 2.8868 & 2.5 & 1577.55 & 104735  & $a/ \Delta y$ = 120 \\
			L9.D.5 & 5 & 3.4641 & 3 & 1577.53 & 104631 & $a/ \Delta z$ = 140  \\
			L9.D.6 & 6 & 4.0415 & 3.5 & 1577.54 & 104642 \\
			L9.D.7 & 7 & 4.6188 & 4 & 1577.52 & 104614 \\
			L9.D.8 & 8 & 5.1962 & 4.5 & 1577.53 & 104616 \\
			\hline
			\textbf{FDFD}  \\
			L9.D.1 & 0.125 & 0.125 & 0.5 & 1579.72 & 105899  & $a/ \Delta x$ = 17.7 \\
			L9.D.2 & 0.185 & 0.185 & 0.56 & 1579.77 & 106598  & $a/ \Delta y$ = 17.7 \\
			L9.D.3 & 0.245 & 0.245 & 0.62 & 1578.72 & 106011  & $a/ \Delta z$ = 17.7 \\
			L9.D.4 & 0.305 & 0.305 & 0.68 & 1578.76 & 106358 \\
			\hline
			\textbf{pFEM}  \\
			L9.D.1 & 0.11 & 0.11 & 0.11 & 1577.64 & 105468 \\
			L9.D.2 & 0.23 & 0.23 & 0.23 & 1577.64 & 105468 \\
			L9.D.3 & 0.46 & 0.46 & 0.46 & 1577.64 & 105467 \\
			L9.D.4 & 0.68 & 0.68 & 0.68 & 1577.64 & 105471 & $p$ = 4 \\
			L9.D.5 & 0.91 & 0.91 & 0.91 & 1577.64 & 105464 & $a/h$ = 2.2 \\
			L9.D.6 & 1.14 & 1.14 & 1.14 & 1577.64 & 105472 \\
			L9.D.7 & 1.71 & 1.71 & 1.71 & 1577.64 & 105473 \\
			L9.D.8 & 2.28 & 2.28 & 2.28 & 1577.64 & 105464 \\
			\hline
			\textbf{sFEM}  \\
			L9.D.1 & 0 & 0 & 0 & 1578.13 & 105449 \\
			L9.D.2 & 0.02 & 0.02 & 0.02 & 1578.13 & 105028 & $p$ = 3  \\
			L9.D.3 & 0.23 & 0.23 & 0.23 & 1578.13 & 105030 & $a/h$ = 3.4 \\
			L9.D.4 & 0.46 & 0.46 & 0.46 & 1578.13 & 102974 \\
			L9.D.5 & 1.14 & 1.14 & 1.14 & 1578.13 & 104620 \\
			\hline
			\textbf{tFEM}  \\
			L9.D.1 & 4 & 4 & 4 & 1579.46 & 104571 & $p$ = 2 \\
			L9.D.2 & 4 & 4 & 6 & 1579.46 & 104390 \\
			L9.D.3 & 4 & 4 & 8 & 1579.46 & 104373 & Min. $a/h $\\
			L9.D.4 & 4 & 4 & 10 & 1579.46 & 104382 & = 1.12 \\
			L9.D.5 & 4 & 4 & 12 & 1579.46 & 104466 \\
			L9.D.6 & 4 & 4 & 14 & 1579.46 & 104442 & Max. $a/h$ \\
			L9.D.7 & 4 & 4 & 16 & 1579.46 & 104519 & = 15.2 \\
			\hline
		\end{tabular}
	\end{threeparttable}	
	\label{Tab:L9_DomainSizeData} 
\end{table}

\section*{Funding}

Teknologi og Produktion, Det Frie Forskningsr\aa d (FTP, DFF) (Sapere Aude LOQIT: DFF-4005-00370); Villum Fonden (VKR Center of Excellence NATEC-II, grant 8692); Deutsche Forschungsgemeinschaft (DFG) (SFB787-B4); Einstein Foundation Berlin (ECMath-OT9).

\end{document}